\documentclass[12pt]{article}
\usepackage{amsfonts,amssymb,amsmath}
\usepackage{epic,eepic}
\usepackage{color}
\usepackage{theorem,cite}
\usepackage{graphics,epsfig}
\definecolor{brique}{rgb}{.9,.2,0}
\definecolor{blvert}{rgb}{0,.8,.85}
\definecolor{vertcl}{rgb}{0,1,.7}
\newcommand\vertcl[1]{\textcolor{vertcl}{#1}}
\newcommand\blvert[1]{\textcolor{blvert}{#1}}
\newcommand\brique[1]{\textcolor{brique}{#1}}
\def\lapth{
\begin{picture}(164,70)(0,-15)\thicklines
\put(0,0){\vertcl{\rule{20pt}{4pt}}}
\put(19,1){\vertcl{\line(1,3){23}}} 
\put(20,1){\vertcl{\line(1,3){23}}} 
\put(21,1){\vertcl{\line(1,3){23}}}
\put(22,1){\vertcl{\line(1,3){23}}}
\put(45,70){\vertcl{\line(1,-3){23}}} 
\put(44,70){\vertcl{\line(1,-3){23}}} 
\put(43,70){\vertcl{\line(1,-3){23}}}
\put(42,70){\vertcl{\line(1,-3){23}}}
\put(2,24){\vertcl{\rule{120pt}{4pt}}}
\put(65,0){\vertcl{\rule{60pt}{4pt}}}
\put(5,37){\Huge{\brique{\textbf{L}}}} 
\put(62,37){\Huge{\brique{\textbf{PTh}}}}
\put(12,-8){\blvert{\rule{92pt}{3.5pt}}}
\put(24,-15){\blvert{\rule{57pt}{3.5pt}}}
\put(36,-22){\blvert{\rule{30pt}{3.5pt}}}
\end{picture}
\raisebox{35pt}{
\begin{minipage}{320pt}\begin{center}
\textbf{Laboratoire d'Annecy-leVieux de Physique
Th\'eorique}\\[4ex]
website: \texttt{http://lappweb.in2p3.fr/lapth-2005/}
\end{center}
\end{minipage}}\\
\vspace{10pt}\quad \hrulefill\\
\vspace{10pt}}

\newtheorem{definition}{Definition}[section]

\newtheorem{proposition}[definition]{Proposition}

\newtheorem{theorem}[definition]{Theorem}

\theorembodyfont{\rmfamily}
\newtheorem{rmk}{Remark}[section]

\numberwithin{equation}{section}

\newcommand{\nonu}{\nonumber \\}

\newcommand{\hs}[1]{\hspace{#1 mm}}

\newcommand{\eps}{\epsilon}

\newcommand{\vph}{\varphi}

\def\cA{{\cal A}}                    \def\cC{{\cal C}}
          \def\cE{{\cal E}}          
                    
                    \def\cL{{\cal L}}
          \def\cN{{\cal N}}          
\def\cP{{\cal P}}                    
          \def\cT{{\cal T}}          \def\cU{{\cal U}}
                    
\def\cY{{\cal Y}}          

\newcommand{\CC}{{\mathbb C}}

\newcommand{\II}{{\mathbb I}}

\newcommand{\ZZ}{{\mathbb Z}}

\newcommand{\wh}[1]{\widehat{#1}}

\newcommand{\mb}[1]{\hs{4}\mbox{#1}\hs{4}}
\newcommand{\qmbox}[1]{{\qquad\mbox{#1}\quad}}
\newcommand{\half}{\frac{1}{2}}

\newcommand{\prf}{\underline{Proof:}\ }
\newcommand{\finprf}{\null \hfill {\rule{5pt}{5pt}}\\[2.1ex]\indent}

\newcommand{\atopn}[2]{\genfrac{}{}{0pt}{}{#1}{#2}}

\newcommand{\enne}{N}
\newcommand{\bs}[1]{{\boldsymbol{#1}}}

\newcommand{\slf}{fundamental $sl(N)$ spin chain }
\newcommand{\sls}{$sl(2)$ spin-s chain }
\newcommand{\alt}{alternating $sl(N)$ spin chain }

\begin{document}
\renewcommand{\thefootnote}{\arabic{footnote}}
\setcounter{footnote}{0}
\newpage
\setcounter{page}{0} \markright{\today\dotfill DRAFT\dotfill }

\pagestyle{empty}

\hspace{-1cm}\lapth

\vfill

\begin{center}

{\Large \textsf{Thermodynamical limit of general $gl(N)$ spin chains: 
\\[1.2ex]
vacuum state and densities}}

\vspace{10mm}

{\large N. Cramp{\'e}$^{ab}$ , L. Frappat$^c$ and
{\'E}. Ragoucy$^c$}
\footnote{E-mails: crampe@sissa.it,
luc.frappat@lapp.in2p3.fr, eric.ragoucy@lapp.in2p3.fr}

\vspace{10mm}
\emph{$^a$ International School for Advanced Studies}

\emph{Via Beirut 2-4, 34014 Trieste, Italy}
\vspace{5mm}\\

\emph{$^b$ Istituto Nazionale di Fisica Nucleare}

\emph{Sezione di Trieste}
\vspace{5mm}\\

\emph{$^c$ Laboratoire d'Annecy-le-Vieux de Physique Th{\'e}orique}

\emph{LAPTH, CNRS, UMR 5108, Universit{\'e} de Savoie}

\emph{B.P. 110, F-74941 Annecy-le-Vieux Cedex, France}\\

\vfill

\end{center}

\vfill
\vfill

\begin{abstract}
We study the vacuum state of spin chains where each site carry an arbitrary
representation. We prove that the string hypothesis, usually used to solve
the Bethe ansatz equations, is valid for representations characterized by
rectangular Young tableaux. In these cases, we obtain the density of the
center of the strings for the vacuum. We work out different examples and,
in particular, the spin chains with periodic array of impurities.
\end{abstract}

\vfill

MSC: 81R50, 17B37 ---
PACS: 02.20.Uw, 03.65.Fd, 75.10.Pq

\vfill

\rightline{LAPTH-1171/07}
\rightline{\texttt{cond-mat/0701207}}
\rightline{January 2007}

\baselineskip=16pt

\newpage
\pagestyle{plain}

\section{Introduction}

Since the seminal work of H.~Bethe \cite{bethe} solving the spin $\half$
Heisenberg spin chain \cite{heisen}, numerous applications of this method
have been appeared to solve more involved models: anisotropic XXZ spin
chain \cite{faddeev, korepin, KoIzBo}; spin 1 chains \cite{ZAFA,MENERI};
alternating spin chains \cite{dewo,martins,abad2,ana}; spin chains with higher
spins \cite{Kul,Tak,Bab,ow,KoIzBo,fafa,KuSu,Bytsko,Tsuboi}; spin chains
with boundaries \cite{Gau,cherednik,sklyanin,alc} or with impurities
\cite{anjo,fuka,YWang}. In the papers \cite{ow,byebye}, the Bethe ansatz
equations (BAE) have been obtained for spin chains based on $gl(N)$ and where
each site may carry a different representation. This generic framework
encompasses the models with higher spin on each site or with impurities and
the alternating spin chains.

In this paper, we are interested in the resolution of these equations
within the thermodynamical limit, but keeping arbitrary the representation
on each site. The first step consists in using the string hypothesis. We
show that the consistency of the BAE implies a restriction on the
representations involved in the spin chain. This restriction is solved by
taking representations characterized by rectangular Young tableaux. For the
other cases, the validity of the string hypothesis remains an open
question. Then, we determine the vacuum state and prove that it is non
degenerate and a trivial representation of the $sl(N)$ symmetry algebra.
This vacuum state is constructed on strings whose lengths and
multiplicities are explicitly given. Finally, the densities of the center
of these strings are computed exactly. Throughout the paper, different
examples are worked out. In particular, we treat in detail the spin chains
with periodic array of impurities for which we compute the energy of the
vacuum state.

The plan of this paper is as follows. In section \ref{sec:gene}, we recall
basic notions and notations on Yangian algebras and spin chains. 
Section \ref{sec:stringhyp} is devoted to the study of the BAE within the
string hypothesis, and the constraints it imposes on the representations
entering the spin chain. The vacuum state is determined in section
\ref{sec:vacuum}, see theorem \ref{thm:vacuum}. Section \ref{sec:tdl} is
concerned with the resolution of the BAE within the thermodynamical limit.
In particular, we compute the densities of the vacuum state, see theorem
\ref{thm:density}.

\section{Generalities\label{sec:gene}}

\subsection{Yangian $\cY(gl(\enne))$\label{sect:ygln}}

We will consider the $gl(\enne)$ invariant $R$ matrices \cite{yang,baxter}
\begin{eqnarray}
  R_{ab}(\lambda) =\II_\enne \otimes \II_\enne +
\frac{i\;\cP_{ab}}{\lambda}
\;,
\label{r}
\end{eqnarray}
where $\cP_{ab}$ is the permutation operator
\begin{equation}
  \label{eq:P12}
  \cP_{ab} = \sum_{i,j=1}^\enne E_{ij} \otimes E_{ji} 
\end{equation}
and $E_{ij}$ are the
elementary matrices with 1 in position $(i,j)$ and 0 elsewhere. 

This $R$ matrix satisfies the {\it Yang--Baxter equation} \cite{mac,yang,baxter,korepin,KoIzBo}
\begin{eqnarray}
  R_{ab}(\lambda_{a}-\lambda_{b})\ R_{ac}(\lambda_{a})\ R_{bc}(\lambda_{b})
  =R_{bc}(\lambda_{b})\ R_{ac}(\lambda_{a})\ 
  R_{ab}(\lambda_{a}-\lambda_{b})\;.
  \label{YBE}
\end{eqnarray}
It can be interpreted physically as a scattering matrix
\cite{zamo, korepin, faddeev} describing the interaction between two
solitons (viewed in this framework as low level excited states in a
thermodynamical limit of a spin chain) that carry the fundamental
representation of $gl(\enne)$.

The Yangian $\cY(gl(\enne))$ \cite{Drinfeld} is the complex associative unital algebra with
the generators $T_{ij}^{(n)}$  ($T_{ij}^{(0)}=\delta_{ij}$), for $1\leq i,j\leq \enne, n\in \ZZ_{\geq
0}$
 subject to the defining relations, called 
FRT exchange relation \cite{FRT},
\begin{eqnarray}
\label{RTT}
R_{ab}(\lambda_a-\lambda_b)\;\cT_a(\lambda_a)\;\cT_b(\lambda_b)=
\cT_b(\lambda_b)\;\cT_a(\lambda_a)\;R_{ab}(\lambda_a-\lambda_b)\;,
\end{eqnarray}
where the generators are gathered in the following matrix (belonging 
to $End(\CC^\enne)\otimes \cY(gl(\enne))[[\lambda^{-1}]]$)
\begin{eqnarray}
\label{def:T}
\cT(\lambda)=\sum_{i,j=1}^\enne E_{ij}\otimes 
T_{ij}(\lambda)
=\sum_{i,j=1}^\enne E_{ij}\otimes 
\sum_{r \geq 0} \frac{1}{\lambda^r}~T_{ij}^{(r)}=
\sum_{r \geq 0} \frac{1}{\lambda^r}~\cT^{(r)}
\;.
\end{eqnarray}
Using the commutation relations (\ref{RTT}), it is easy to show that 
$\cT^{(1)}$ generates a $gl(\enne)$ algebra.\\

\subsection{Algebraic transfer matrix}

In the following, in order to construct spin chains, it will be
necessary to deal with the tensor product of $L$ copies of the
Yangian. For $1\leq i \leq L$, we denote by $\cL_{ai}(\lambda)\in End
(\CC^\enne) \otimes \cY(gl(\enne)) $ one copy of the Yangian which
acts non trivially on the $i^{th}$ space only. The space $a$, 
always isomorphic to $End(\CC^\enne)$ in the present paper, is called
 auxiliary space whereas the space $i$ is called  quantum
space. Mimicking the notation (\ref{def:T}) used for  $\cT(\lambda)$, we
will consider $\cL_{ai}(\lambda)$ as an $\enne\times\enne$ matrix
whose entries are operators acting on the quantum space $i$. These
operators will be expanded as series in $\lambda$, so that we have
\begin{eqnarray}
\cL_{ai}(\lambda)=\sum_{p,q=1}^\enne E_{pq}\otimes 
L_{pq}(\lambda)
=\sum_{p,q=1}^\enne E_{pq}\otimes 
\sum_{r \geq 0} \frac{1}{\lambda^r}~L_{pq}^{(r)}=
\sum_{r \geq 0} \frac{1}{\lambda^r}~\cL_{ai}^{(r)}
\;.
\end{eqnarray}
Obviously, $\cL_{ai}(\lambda)$ satisfies the defining relations
of the Yangian
\begin{eqnarray}
\label{Rll}
R_{ab}(\lambda_a-\lambda_b)\;\cL_{ai}(\lambda_a)\;\cL_{bi}(\lambda_b)=
\cL_{bi}(\lambda_b)\;\cL_{ai}(\lambda_a)\;R_{ab}(\lambda_a-\lambda_b)\;.
\end{eqnarray}
Let us stress that the matrix $\cL_{ai}(\lambda)$ is local, i.e. it 
contains only
the $i^{th}$ copy of the Yangian. On the contrary, 
one constructs a non-local algebraic object, the monodromy
matrix
\begin{align}
\label{mono}
\cT_a(\lambda)&=
\cL_{a1}(\lambda)\;\cL_{a2}(\lambda)\;\dots\;\cL_{aL}(\lambda)\in End
(\CC^\enne) \otimes \left(\cY(gl(\enne))\right)^{\otimes L} \;.
\end{align}
Let us remark that the quantum spaces are omitted in the LHS of
(\ref{mono}), as usual in the notation of the monodromy matrix. The
entries of the monodromy matrix $\cT_a(\lambda)$ are given by
\begin{align}
\label{mono2}
T_{ij}(\lambda)&=\sum_{k_1,\dots,\,k_{L-1}=1}^\enne
L_{ik_1}(\lambda)\;\otimes\;L_{k_1k_2}(\lambda)\;\otimes\;
\dots\;\otimes\;L_{k_{L-1}j}(\lambda)\;.
\end{align}
and satisfies the defining relations of the Yangian
\begin{eqnarray}
\label{Rtt}
R_{ab}(\lambda_a-\lambda_b)\;\cT_{a}(\lambda_a)\;\cT_{b}(\lambda_b)=
\cT_{b}(\lambda_b)\;\cT_{a}(\lambda_a)\;R_{ab}(\lambda_a-\lambda_b)\;.
\end{eqnarray}
Now, we can introduce the main object for the study of spin
chains, i.e. the transfer matrix 
\begin{eqnarray}
\label{transf}
t(\lambda)
=tr_{a}\left(\cT_{a}(\lambda)\right)
=\sum_{i=1}^\enne T_{ii}(\lambda)\;.
\end{eqnarray}
Equation (\ref{Rtt}) immediately implies
\begin{eqnarray}
\label{comt}
[\;t(\lambda)\;,\;t(\mu)\;]=0
\end{eqnarray}
which will guarantee the integrability of the models.

Let us remark that, at that point, the monodromy and transfer 
matrices are algebraic objects (in $\left(\cY(gl(\enne))\right)^{\otimes 
L}$), and, as such, play the r{\^o}le of generating functions for the 
construction of monodromy and transfer  matrices as they are usually 
introduced in spin chain models. The latter will be constructed from 
the former using representations of the Yangian, as it will be done 
below.

\subsection{Representations \label{reps}}

In order to construct representations of $\cY(gl(\enne))$, the
following algebra homomorphism (called evaluation map, see e.g. \cite{MNO})
from $\cY(gl(\enne))$ to $\cU(gl(\enne))$
(universal enveloping algebra of $gl(\enne)$)
will be used
\begin{eqnarray}
\label{eval}
L_{pq}(\lambda)\longmapsto \delta_{pq}+\frac{i\;e_{qp}}{\lambda}\;,
\end{eqnarray}
where $e_{pq}$ ($1\leq p,q \leq \enne$) are the abstract generators of the 
Lie algebra $gl(\enne)$. They satisfy the following commutation relations
\begin{equation}
[e_{pq},e_{kl}]=\delta_{qk}\;e_{pl}-\delta_{lp}\; e_{kq} \;.
\end{equation}
Spin chain models will be obtained through the 
evaluation of the algebraic monodromy and transfer matrices in 
Yangian representations. We thus present here some basic results on 
the classification of finite-dimensional irreducible representations 
of $\cY(gl(\enne))$.

\subsubsection{Evaluation representations}

Keeping in mind the forthcoming spin chains interpretation, we choose for
each local $\cY(gl(\enne))$ algebra an irreducible finite-dimensional
evaluation representation. 

We start with a finite-dimensional irreducible representation of
$gl(\enne)$, $M({\bs \varpi})$, with highest weight ${\bs \varpi} =
(\varpi^{(1)},\dots,\varpi^{(N)})$ and associated to the highest weight vector
$v$. This highest weight vector obeys
\begin{eqnarray}
&&e_{kj}\;v=0 \qmbox{,} 1\leq k<j \leq \enne  \\
&&e_{kk}\;v=\varpi^{(k)}\; v \qmbox{,}  1\leq k \leq \enne \;,
\end{eqnarray}
where $\varpi^{(1)},\dots,\varpi^{(N)}$ are integers with
$\varpi^{(k+1)}\leq \varpi^{(k)}$ (these constraints on the parameters
$\varpi^{(k)}$ are criteria so that the representation be
finite-dimensional and irreductible). 

The evaluation representation $M_\lambda({\bs \varpi})$ of 
$\cY(gl(\enne))$ is built from $M({\bs \varpi})$ and
follows from the homomorphism (\ref{eval}), according to
\begin{eqnarray}
&&L_{jk}(\lambda)\;v=0 \qmbox{,} 1\leq k<j \leq \enne  \\
&&L_{kk}(\lambda)\;v=\left(1+\frac{i\;\varpi^{(k)}}{\lambda}\right)\; v 
\qmbox{,}  1\leq k \leq \enne \;.
\end{eqnarray}

The representation $M_\lambda((1,0,\dots,0))$, associated to the $gl(\enne)$
fundamental representation, of $\cL(\lambda)$ 
provides the $R$ matrix (\ref{r}).

\subsubsection{Representations of the monodromy matrix \label{decadix}}

The evaluation representations of $\cL(\lambda)$ allow us to build a
representation of the monodromy matrix. Indeed, evaluating each of the
local $\cL_{a,\ell}(\lambda)$ in a representation
$M_{\lambda+i\theta_{\ell}}(\bs{\varpi_{\ell}})$ for $1\leq \ell \leq L$, the tensor
product built on
\begin{eqnarray}
\label{tensorp}
M_{\lambda+i\theta_{1}}(\bs{\varpi_{1}}) \otimes \dots \otimes
M_{\lambda+i\theta_{L}}(\bs{\varpi_{L}})
\end{eqnarray}
provides, via (\ref{mono2}), a finite-dimensional representation  for 
$\cT(\lambda)$. 
The shifts $i\theta_{\ell}$ entering in the definition of the
representations are called inhomogeneity parameters.
Denoting by $v_{\ell}$ the highest weight vector associated to
$\boldsymbol{\varpi_{\ell}}=(\varpi^{(1)}_{\ell},\dots,\varpi^{(N)}_{\ell})$, 
the vector 
\begin{equation}
    v^+=v_1 \otimes \dots \otimes v_{L}
\label{v+}
\end{equation}
is the highest weight vector of the representation (\ref{tensorp})
i.e.\footnote{We have renormalized the generators in such a way that
the represented monodromy matrix is analytical in $\lambda$.}
\begin{eqnarray}
  && T_{jk}(\lambda)\;v^+=0 \qmbox{,} 1\leq k<j \leq N  \\
  && T_{jj}(\lambda)\;v^+=\prod_{\ell=1}^L
\left(\lambda+i\theta_{\ell}+i\;\varpi^{(j)}_{\ell}\right)\;v^+ \qmbox{,} 
  1\leq j \leq N \;.
\end{eqnarray}
In the following, we use the following notation, for $1\leq j \leq
N$ 
\begin{eqnarray}
P_j(\lambda)=\prod_{\ell=1}^L
\left(\lambda+i\theta_{\ell}+i\;\varpi^{(j)}_{\ell}\right)\;.\label{drinP}
\end{eqnarray}
These polynomials, related to Drinfel'd polynomials, are usually introduced
 to classify the representations of Yangians.

\subsection{Pseudo-vacuum and dressing functions}

We now compute the eigenvalues of the transfer matrix $
t(\lambda)$. As a  by-product, they will provide the Hamiltonian
eigenvalues. 

The highest weight vector (\ref{v+}) is obviously an eigenvector of the
transfer matrix. Indeed, one gets
\begin{eqnarray}
 t(\lambda)\;v^+=\sum_{k=1}^N ~ T_{kk}(\lambda)\;v^+
=\Lambda^0(\lambda)\;v^+
\end{eqnarray}
with
\begin{eqnarray}
\Lambda^0(\lambda)=\sum_{k=1}^N ~
P_k(\lambda)\;.
\end{eqnarray}
Note that $\Lambda^0(\lambda)$ is analytical. In the context of the spin
chains, the highest weight vector $v^+$ is called the pseudo-vacuum. The
next step consists in the ansatz itself which provides all the
eigenvalues of $ t(u)$ from $\Lambda^0(\lambda)$.

We make the following assumption for the structure of all the
eigenvalues of $ t(u)$ 
\begin{eqnarray}
\Lambda(\lambda)=\sum_{k=1}^N ~
P_k(\lambda)\;
D_k(\lambda)\;,\label{ClosedEigen}
\end{eqnarray}
where $D_k(\lambda)$, the so-called dressing functions, have  been
determined in \cite{byebye}. They are rational functions of the 
form
\begin{eqnarray}
D_k(\lambda)=\prod_{n=1}^{M^{(k-1)}}
\frac{\lambda-\lambda_n^{(k-1)}+\frac{i\;(k+1)}{2}}
{\lambda-\lambda_n^{(k-1)}+\frac{i\;(k-1)}{2}}
~~\prod_{n=1}^{M^{(k)}}
\frac{\lambda-\lambda_n^{(k)}+\frac{i\;(k-2)}{2}}
{\lambda-\lambda_n^{(k)}+\frac{i\;k}{2}}\;.
\label{dressingClosed}
\end{eqnarray}
where $M^{(0)}=M^{(N)}=0$. The parameters $M^{(k)}\in\ZZ_{+}$, 
$k=1,\ldots,N$ and $\lambda_{n}^{(k)}\in\CC$,
$n=1,\ldots,M^{(k)}$, $k=1,\ldots,N$ are determined through the
celebrated Bethe Ansatz equations (see below). 
One can also relate the parameters $M^{(k)}$, $k=1,\ldots,N$,
 to the $T^{(1)}_{jj}$ eigenvalues, according to
\begin{equation}
T^{(1)}_{jj}\,w = \Big(M^{(j-1)} -M^{(j)}+\sum_{\ell=1}^L \theta_{\ell}+\varpi^{(j)}_{\ell}
\Big)\,w
\end{equation}
where $w$ is the transfer matrix eigenvector with eigenvalue (\ref{ClosedEigen}).

When considering the $sl(N)$ generators $ T^{(1)}_{jj}-
T^{(1)}_{j+1,j+1}$, the eigenvalues read
\begin{equation}
   S_{j}= M^{(j-1)}+M^{(j+1)} -2\,M^{(j)}+\sum_{\ell=1}^L a^{(j)}_{\ell}
\label{eq:spin-M}
\end{equation}
 where $a^{(j)}_{\ell}=\varpi^{(j)}_{\ell}-\varpi^{(j+1)}_{\ell}\in
 \ZZ_{\geq0}$. 
 
 Remark that for $sl(2)$, $S_{j}$ corresponds to twice the `usual'
 spin.

\section{String hypothesis\label{sec:stringhyp}}

\subsection{BAE for the center of strings}

The following set of Bethe ansatz equations, for
$j=1,\ldots,N-1$ and $k=1,\ldots,M^{(j)}$ has been obtained in
\cite{KulResh,ow} (see also \cite{byebye}):
\begin{eqnarray}
\label{eq:bethe} \frac{P_{j}(\lambda_{k}^{(j)}-\frac{ij}{
2})}{P_{j+1}(\lambda_{k}^{(j)}-\frac{ij}{2})} =
\prod_{\ell=1}^{M^{(j-1)}}
   e_{-1}(\lambda_{k}^{(j)} -\lambda_{\ell}^{(j-1)}) \
   \prod_{\atopn{\ell=1}{\ell\neq k}}^{M^{(j)}}
   e_{2}(\lambda_{k}^{(j)} -\lambda_{\ell}^{(j)})\
   \prod_{\ell=1}^{M^{(j+1)}}
   e_{-1}(\lambda_{k}^{(j)} -\lambda_{\ell}^{(j+1)})\quad
\end{eqnarray}
where we defined
\begin{equation}
  e_{p}(\lambda) =
  \frac{\lambda+\frac{ip}{2}}{\lambda-\frac{ip}{2}}\;.
\end{equation}
Knowing that
$a^{(j)}_{\ell}=\varpi^{(j)}_{\ell}-\varpi^{(j+1)}_{\ell}\in \ZZ_+$, the
L.H.S. of (\ref{eq:bethe}) can be written as
\begin{eqnarray}
\frac{P_{j}(\lambda_{k}^{(j)}-\frac{ij}{
2})}{P_{j+1}(\lambda_{k}^{(j)}-\frac{ij}{ 2})}=\prod_{\ell=1}^L
e_{a^{(j)}_{\ell}}\left(\lambda_{k}^{(j)}+i\theta_\ell
+i\frac{\varpi^{(j)}_{\ell}+\varpi^{(j+1)}_{\ell}-j}{2}\right)
\end{eqnarray}

We postulate the string hypothesis which states that all the
roots $\{\lambda_{1}^{(j)},\dots, \lambda_{M^{(j)}}^{(j)}\}$ are
gathered into $\nu^{(j)}_{m}$ strings of length $2m+1$ $(m\in
\half{\ZZ_+})$ of the following form
\begin{equation}
\lambda^{(j)}_{m,k,\alpha}=\lambda^{(j)}_{m,k}+i\,\alpha
\,,\quad \alpha=-m,-m+1,\dots, m
\end{equation}
where $k=1,\dots,\nu^{(j)}_{m}$ and $\lambda^{(j)}_{m,k}$, the center of
the string, is real. 

The string hypothesis is sustained by numerical
calculations, done for a very large number of types of strings. 
One has to keep in mind that it is valid only 
in the thermodynamic limit $L\rightarrow
+\infty$, which will be always the case that we will consider below.
Under this hypothesis, we have
\begin{equation}
\sum_{m\in \half\ZZ_+}(2m+1)\,\nu^{(j)}_{m}=M^{(j)}\;.
\end{equation}
Because of the vanishing of $M^{(0)}$ and $M^{(N)}$, 
the parameters $\nu^{(0)}_{m}$ and $\nu^{(N)}_{m}$ vanish also.
We can deduce from (\ref{eq:bethe}) a set of equations for the
centers of the strings, for $j=1,\dots, N-1$, $m\in\half \ZZ_+$,
$k=1,\dots ,\nu^{(j)}_{m}$. Indeed, considering products of BAE's
coming from a string, we get: 
\begin{eqnarray}
\label{eq:bethe-center} &&\prod_{\ell=1}^L
E_{a^{(j)}_{\ell}}^{(m)}\left(\lambda_{m,k}^{(j)}+i\theta_\ell
+i\frac{\varpi^{(j)}_{\ell}+\varpi^{(j+1)}_{\ell}-j}{2}\right)\\
&&= \prod_{p\in\half\ZZ_+}\left(
\prod_{\ell=1}^{\nu^{(j-1)}_{p}}E_{-1}^{(p,m)}(\lambda^{(j)}_{m,k}-\lambda^{(j-1)}_{p,\ell})
\prod_{\atopn{\ell=1}{(p,\ell)\neq(m,k)}}^{\nu^{(j)}_{p}}E_{2}^{(p,m)}(\lambda^{(j)}_{m,k}-\lambda^{(j)}_{p,\ell})
\prod_{\ell=1}^{\nu^{(j+1)}_{p}}E_{-1}^{(p,m)}(\lambda^{(j)}_{m,k}-\lambda^{(j+1)}_{p,\ell})
\right)\nonumber
\end{eqnarray}
where $(p,\ell)\neq(m,k)$ means that when $p=m$, one has to discard
the term corresponding to $\ell=k$ in the product. We have introduced
\begin{eqnarray}
E_{p}^{(m)}(\lambda) &=& 
e_{p-2m}(\lambda)\,e_{p-2m+2}(\lambda)\,\ldots e_{p+2m}(\lambda)\\
E_{2}^{(m,n)}(\lambda) &=& 
\begin{cases} 
e_{4m+2}(\lambda)\,\Big(e_{4m}(\lambda)\,
e_{4m-2}(\lambda)\,\ldots
e_{2}(\lambda)\Big)^2 &\mbox{ if }m= n\\
e_{2(m+n+1)}(\lambda)\,\Big(e_{2(m+n)}(\lambda)\,e_{2(m+n-1)}(\lambda)\,\ldots
e_{2\vert m-n\vert +2}(\lambda)\Big)^2\,e_{2\vert m-n\vert}(\lambda)
&\mbox{ if }m\neq n\end{cases}
\quad\quad\\
E_{-1}^{(m,n)}(\lambda) &=& 
\Big(e_{2\vert m-n\vert+1}(\lambda)\,e_{2\vert m-n\vert+3}(\lambda)
\ldots e_{2(m+n)+1}(\lambda)\Big)^{-1}
\end{eqnarray}

\subsection{String hypothesis and constraint on the representations}

{F}rom the string hypothesis, the $\lambda^{(j)}_{m,k}$ parameters are
real, so that the r.h.s. of equation (\ref{eq:bethe-center}) is of
modulus one. Then, the l.h.s. of this equation must also be of modulus
one. This condition implies some constraints on the type of 
representations that can be on the chain:
\begin{proposition}
The BAE of the string centers are consistent if and only if
the representations entering the 
spin chain solve the equations
\begin{equation}
\sum_{\ell=1}^{L}  
\left(\gamma^{(j)}_{\ell}+\frac{a^{(j)}_{\ell}}{2}\right)^{2p}\, 
=\,\sum_{\ell=1}^{L}  
\left(\gamma^{(j)}_{\ell}-\frac{a^{(j)}_{\ell}}{2}\right)^{2p}
\,,\qquad 0\leq p\leq L\,,\ 1\leq j\leq N
\label{eq:CNSstring}
\end{equation}
where we have introduced
\begin{equation}
\gamma^{(j)}_{\ell}=\theta_\ell
+\frac{\varpi^{(j)}_{\ell}+\varpi^{(j+1)}_{\ell}-j}{2}\,.
\label{eq:gamma}
\end{equation}
\end{proposition}
\prf
Demanding the l.h.s. of (\ref{eq:bethe-center}) to be of modulus 1, is equivalent to ask its 
conjugate to be its inverse, that is
\begin{equation}
\prod_{\ell=1}^{L} \prod_{q=-m}^{m} 
e_{a^{(j)}_{\ell}+2q}\Big(\lambda+i\,\gamma^{(j)}_{\ell}\Big)\ 
e_{-a^{(j)}_{\ell}-2q}\Big(\lambda-i\,\gamma^{(j)}_{\ell}\Big)=1
\end{equation}
where for simplicity we abrievated $\lambda^{(j)}_{m,k}$ into $\lambda$.
These equations must be fulfilled whatever the type of strings, i.e. 
for all $m$. Then, they are equivalent to
\begin{eqnarray*}
&&\prod_{\ell=1}^{L} 
e_{a^{(j)}_{\ell}}\Big(\lambda+i\,\gamma^{(j)}_{\ell}\Big)\ 
e_{-a^{(j)}_{\ell}}\Big(\lambda-i\,\gamma^{(j)}_{\ell}\Big)=1\,,\ \
\\
&&\prod_{\ell=1}^{L} 
e_{a^{(j)}_{\ell}+n}\Big(\lambda+i\,\gamma^{(j)}_{\ell}\Big)\ 
e_{a^{(j)}_{\ell}-n}\Big(\lambda+i\,\gamma^{(j)}_{\ell}\Big)\ 
e_{-a^{(j)}_{\ell}-n}\Big(\lambda-i\,\gamma^{(j)}_{\ell}\Big)\ 
e_{-a^{(j)}_{\ell}+n}\Big(\lambda-i\,\gamma^{(j)}_{\ell}\Big)=1\,,\ \
\forall\ n\in\ZZ_{>0}\quad\ 
\end{eqnarray*}
which leads to
\begin{eqnarray*}
\prod_{\ell=1}^{L}
\left(\lambda_{+}^2+\big(\gamma^{(j)}_{\ell}+\frac{a^{(j)}_{\ell}}{2}\big)^2\right)
\left(\lambda_{-}^2+\big(\gamma^{(j)}_{\ell}+\frac{a^{(j)}_{\ell}}{2}\big)^2\right)
= \prod_{\ell=1}^{L}
\left(\lambda_{+}^2+\big(\gamma^{(j)}_{\ell}-\frac{a^{(j)}_{\ell}}{2}\big)^2\right)
\left(\lambda_{-}^2+\big(\gamma^{(j)}_{\ell}-\frac{a^{(j)}_{\ell}}{2}\big)^2\right)\quad
\end{eqnarray*}
where $\lambda_{\pm}=\lambda\pm i\frac{n}{2}$. Since this equation 
must be satisfied for all $\lambda$ and $n$, it must be obeyed for 
all $\lambda_{\pm}$. Then, it is equivalent to
\begin{eqnarray}
&&\prod_{\ell=1}^{L}
\left(\lambda^2+\Big(\gamma^{(j)}_{\ell}+\frac{a^{(j)}_{\ell}}{2}\Big)^2\right)\,=
\prod_{\ell=1}^{L}
\left(\lambda^2+\Big(\gamma^{(j)}_{\ell}-\frac{a^{(j)}_{\ell}}{2}\Big)^2\right)
\,,\quad\forall\ \lambda\,.
\end{eqnarray}
Annihilating the coefficients of this polynomial in $\lambda^{2}$ 
gives (after some algebra) equations (\ref{eq:CNSstring}).\finprf

One has to keep in mind that the string hypothesis is valid only in 
the $L\to\infty$ limit. Then, the equations (\ref{eq:CNSstring}) must 
be solved for any value of $L$. These equations can be alternatively 
written as 
\begin{equation}
\sum_{\ell=1}^L \left\{
\gamma^{(j)}_{\ell}\,a^{(j)}_{\ell}\ \sum_{n=0}^{p-1}
\left(\gamma^{(j)}_{\ell}+\frac{a^{(j)}_{\ell}}{2}\right)^{2n}\, 
\,\left(\gamma^{(j)}_{\ell}-\frac{a^{(j)}_{\ell}}{2}\right)^{2p-2n-2}
\right\}\,=\,0\,.
\label{eq:factor-const}
\end{equation}
Hence, a sufficient condition to solve these constraints is to ask
$a^{(j)}_{\ell}\gamma^{(j)}_{\ell}=0$ for all values of $\ell$ and $j$, that 
is
\begin{equation}
\theta_\ell
+\frac{\varpi^{(j)}_{\ell}+\varpi^{(j+1)}_{\ell}-j}{2}=0\mb{or}a^{(j)}_{\ell}=0
\,,\qquad \forall
\ell=1,\ldots,L\,,\ \forall j=1,\ldots,N-1. \label{eq:phase}
\end{equation}
Assuming
moreover that no site carries a trivial representation, i.e. 
$\forall \ell$, there exists at least one $j$ such that 
$a^{(j)}_{\ell}\neq 0$, the above
equation is equivalent to the requirement that for all site $\ell$ this
$j$ is unique. We will call $j_{\ell}$ this unique $j$ and $a_{\ell}$
the unique non-vanishing $a^{(j)}_{\ell}$.

\begin{figure}[ht]
\begin{displaymath}
  \raisebox{22pt}{$j_\ell \left\{
  \rule{0pt}{30pt}
  \right.$}
  \begin{picture}(60,60)
    \put(0,30){\line(1,0){20}}\put(50,30){\line(1,0){10}}
    \put(0,50){\line(1,0){20}}\put(50,50){\line(1,0){10}}
    \put(0,40){\line(1,0){20}}\put(50,40){\line(1,0){10}}
    \put(0,50){\line(0,-1){20}}\put(50,50){\line(0,-1){20}}
    \put(10,50){\line(0,-1){20}}\put(60,50){\line(0,-1){20}}
    \put(20,50){\line(0,-1){20}}
    \put(0,10){\line(1,0){20}}\put(50,10){\line(1,0){10}}
    \put(0,0){\line(1,0){20}}\put(50,0){\line(1,0){10}}
    \put(0,10){\line(0,-1){10}}\put(50,10){\line(0,-1){10}}
    \put(10,10){\line(0,-1){10}}\put(60,10){\line(0,-1){10}}
    \put(20,10){\line(0,-1){10}}
    \multiput(20,50)(4,0){8}{\line(1,0){2}}
    \multiput(20,40)(4,0){8}{\line(1,0){2}}
    \multiput(20,30)(4,0){8}{\line(1,0){2}}
    \multiput(20,10)(4,0){8}{\line(1,0){2}}
    \multiput(20,0)(4,0){8}{\line(1,0){2}}
    \multiput(0,10)(0,4){5}{\line(0,1){2}}
    \multiput(10,10)(0,4){5}{\line(0,1){2}}
    \multiput(20,10)(0,4){5}{\line(0,1){2}}
    \multiput(50,10)(0,4){5}{\line(0,1){2}}
    \multiput(60,10)(0,4){5}{\line(0,1){2}}
    \put(0,-3){$\underbrace{~~~~~~~~~~~~~~~}_{\displaystyle a_\ell}$}
  \end{picture}
\end{displaymath}
    \caption{Young tableau of $sl(N)$\label{Youngtab}}
\end{figure}
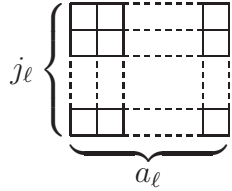

The corresponding $gl(N)$ and $sl(N)$ highest weights  read
\begin{eqnarray}
gl(N) &:& \Big(
\underbrace{\varpi^{(1)}_{\ell},\varpi^{(1)}_{\ell},\ldots,\varpi^{(1)}_{\ell}}_{j_{\ell}},
\underbrace{\varpi^{(1)}_{\ell}-a_{\ell},\ldots,\varpi^{(1)}_{\ell}-a_{\ell}}_{N-j_{\ell}}\big)
\label{eq:glN-weight}\\
sl(N)
&:&(\underbrace{0,\ldots,0}_{j_{\ell}-1},a_{\ell},\underbrace{0,\ldots,0}_{N-1-j_{\ell}})
\end{eqnarray}
and the inhomogeneity parameters are related to the $gl(1)$ eigenvalue
$\varpi_\ell^{(1)}$ through:
\begin{equation}
\theta_\ell=\frac{j_\ell+a_{\ell}}{2}-\varpi^{(1)}_{\ell}\,,\qquad \forall
\ell=1,\ldots,L\;. \label{eq:inhomg}
\end{equation}
Once the $sl(N)$ representation has been chosen for a site, the value 
of  $\varpi_\ell^{(1)}$ is still free and can be used to fix the inhomogeneity parameter 
to an arbitrary value (see examples below and in section \ref{sec:periodic-array}). 
In terms of $sl(N)$ representations, this constraint 
means that we will consider only representation associated with rectangular 
Young tableau, see figure \ref{Youngtab}.

{From} the above discussion, one could wonder whether other kinds of spin 
chains (these which do not solve the equations (\ref{eq:CNSstring}))
 are well-defined. In fact, these
spin chain models do exist and are integrable: the
constraint (\ref{eq:CNSstring}) is just a consequence of the string
hypothesis, and is not related to the existence of the model.

{From} now on, we will restrict ourself to the case of rectangular Young 
tableaux, so as to use the string hypothesis. It should be clear that 
this is a restriction, and that other types of spin chains allowing 
string solutions do exist. However, it does not seem to exist general 
classes as the one given by (\ref{eq:glN-weight}). We will come back on this point in 
 section \ref{sec:L0rep}.
 
Let us also remark that the above restriction permits to consider 
all the representation of $sl(2)$ and all the
fundamental representations of $sl(N)$. It also allows to have different
representations on each site of the chain.

\paragraph{Examples}
Throughout the paper, three examples will be taken to illustrate the general formulae. 
We will consider:
\begin{enumerate}
\item \textit{A \slf}\!\!, that  is a spin chain where
all the sites carry a $sl(N)$ fundamental representation i.e. $j_{\ell}=1=a_{\ell}$,
$\forall \ell$.  As explained above, the inhomogeneity parameters on each site 
may be chosen arbitrarily thanks to the $gl(1)$ parameters $\varpi_\ell^{(1)}$. 
For example, we recover the case where all the inhomogeneity parameters vanish 
if $\varpi_\ell^{(1)}=1$ i.e. the corresponding $gl(N)$ highest weight is
\begin{equation}
\Big(
\underbrace{1,\ldots,1}_{j_{\ell}},
\underbrace{0,\ldots,0}_{N-j_{\ell}}\big)
\end{equation}
\item \textit{A \sls}\!\!, which is a spin chain where all the sites carry a spin $s$ representation 
of $sl(2)$ i.e. $a_{\ell}=2s, j_\ell=1$, $\forall \ell$.
 
\item \textit{An \alt}\!\!, where the sites $2\ell+1$ 
carry an $sl(N)$ representation defined by $(j_{1},a_{1})$ and the sites $2\ell$ 
carry a representation $(j_{2},a_{2})$.
\end{enumerate}

\subsection{Phases of the BAE}

Due to the constraint on the representations, the equations (\ref{eq:bethe-center})
reduce to equations on  phases. To express them, we 
introduce
\begin{eqnarray}
 \vph_{p}(\lambda) &=& 2\,\mbox{Arctan}\left(\frac{2\lambda}{p}\right)\,,\
 p\in\ZZ\,,\ p\neq0\mb{and}
 \vph_{0}(\lambda) = 0
\end{eqnarray}
which obeys
$$
e_{p}(\lambda) = e^{i\,\vph_{p}(\lambda)}\ ,\quad \forall\ p\in\ZZ\,.
$$
Accordingly, we define:
\begin{eqnarray}
 \Phi_{p}^{(m)}(\lambda)&=& \sum_{\alpha=-m}^m \vph_{p+2\alpha}(\lambda)=\sum_{\alpha=\vert \frac{p}{2}-m\vert+1}^{\frac{p}{2}+m}\vph_{2\alpha}(\lambda)+\theta(p>2m)\vph_{p-2m}(\lambda)\,,\
 p\in\ZZ_{+}\ ,\ m\in\half\ZZ_{+}\qquad\\
 \Phi_{2}^{(m,n)}(\lambda) &=& 
\begin{cases} \displaystyle
\vph_{4m+2}(\lambda)\,+2\,\sum_{\alpha=1}^{2m} \vph_{2\alpha}(\lambda)\,
 &\mbox{ if }m= n\\[1.2ex]
\displaystyle \vph_{2m+2n+2}(\lambda)\,+
\,\vph_{2\vert m-n\vert}(\lambda)\,+2\sum_{\alpha=\vert m-n\vert+1}^{m+n}
\vph_{2\alpha}(\lambda)
&\mbox{ if }m\neq n\end{cases}
\quad\quad\\
\Phi_{-1}^{(m,n)}(\lambda) &=& -\sum_{\alpha=\vert m-n\vert}^{m+n}
\vph_{2\alpha+1}(\lambda)
\end{eqnarray}
where we have introduced the step function $\theta(s> r)=1$ if $s-r>0$ and 0 otherwise.
Then, the BAE can be rewritten as
\begin{eqnarray}
\label{eq:bethe-argumt} &&\sum_{\ell=1}^L
\delta_{j,j_{\ell}}\,\Phi_{a_{\ell}}^{(m)}\left(\lambda_{m,k}^{(j)}\right) -
2\pi\,Q_{m,k}^{(j)}=
\label{eq:phasBAE}\\
&& \sum_{p\in\half\ZZ_+}\left(
\sum_{\ell=1}^{\nu^{(j-1)}_{p}}
\Phi_{-1}^{(p,m)}(\lambda^{(j)}_{m,k}-\lambda^{(j-1)}_{p,\ell})+
\sum_{\atopn{\ell=1}{(p,\ell)\neq(m,k)}}^{\nu^{(j)}_{p}}
\Phi_{2}^{(p,m)}(\lambda^{(j)}_{m,k}-\lambda^{(j)}_{p,\ell})+
\sum_{\ell=1}^{\nu^{(j+1)}_{p}}
\Phi_{-1}^{(p,m)}(\lambda^{(j)}_{m,k}-\lambda^{(j+1)}_{p,\ell})
\right)\nonumber
\end{eqnarray}
where the parameters $Q_{m,k}^{(j)}$ are integer or half-integer,
depending on
the type of strings.

We shall now use the monotony hypothesis, which states that
$Q_{m,k}^{(j)}$ increases with the Bethe's root $\lambda_{m,k}^{(j)}$.
This hypothesis is also confirmed by numerical calculations. As an
illustration, we solved numerically the BAE (\ref{eq:phasBAE})
for the vacuum state of an alternating $gl(2)$ spin
chain with spins $s=1/2$ and $s=3/2$, and a number $L=128$ of sites. We
then plotted the vacuum density computed in the thermodynamical limit 
(see section \ref{sec:tdl}) and
its discrete version\footnote{For $gl(2)$, all the non-zero
densities have the same curve, so that one needs not to distinguish the 
strings of different length, see below. \label{foot:length}}
\begin{equation}
\bar\sigma(\lambda_{k}) =\frac{2}{L\,(\lambda_{k+1}-\lambda_{k})}\ ,
\end{equation}
where $\lambda_{k}$ are the ordered solutions.
Since the densities are computed thanks to the monotony hypothesis,
the matching between them and their discrete analogues confirms the
hypothesis, at least for the vacuum state (see figure
\ref{fig:numeric}). 

This hypothesis allows us to get the bounds on $Q_{m,k}^{(j)}$
sending $\lambda_{m,k}^{(j)}$ to $\pm\infty$. A direct calculation
shows that
\begin{eqnarray}
&&    \Phi_{2}^{(p,m)}(\pm\infty) =
    \pm\pi\,\Big(4\,\min(m,p)+2-\delta_{m,p}\Big)\quad;\quad
\Phi_{-1}^{(p,m)}(\pm\infty) = \mp\pi\,\Big(2\,\min(m,p)+1\Big)\qquad\\
&&\Phi_{a}^{(m)}(\pm\infty) = \pm\pi\,\min(2m+1,a)
\end{eqnarray}
and for $j=1,\ldots,N$, $m\in\half\ZZ_{+}$:
\begin{eqnarray} 
    Q^{(j)}_{m,\pm\infty} = \pm\half\Big(\nu^{(j)}_{m}+4m+1
  +\sum_{\ell=1}^{L}\delta_{j,j_{\ell}}\,\min(2m+1, a_{\ell})
 -\sum_{n\in\half\ZZ_{+}}
\min(2m+1,2n+1)\, w_{n}^{(j)}\Big)\quad
\end{eqnarray}
where we have introduced 
\begin{equation}
w_{m}^{(j)} = 2 \nu^{(j)}_{m}-\nu^{(j-1)}_{m}-\nu^{(j+1)}_{m}
\label{eq:defw}
\end{equation}
with the convention $w_{m}^{(0)} =w_{m}^{(N+1)} =0$, $\forall
m\in\half\ZZ_{+}$.

The reached bounds are deduced from the limiting values
$Q^{(j)}_{m,\pm\infty}$ shifting them by the length of a string:
\begin{eqnarray}
   Q^{(j)}_{m,\min} = Q^{(j)}_{m,-\infty}+2m+1 \mb{and}
   Q^{(j)}_{m,\max}=Q^{(j)}_{m,+\infty}-(2m+1)
   \label{eq:Qminmax}
\end{eqnarray}
 Since we assumed the monotony hypothesis, all the $Q^{(j)}_{m,k}$,
 for $j$ and $m$ fixed, are different one from each other. Then, the
 above bounds, which indicate the possible values for the $Q^{(j)}_{m,k}$,
 also show, through combinatorics, the maximal number of possible
 states. It is thus natural to introduce:
\begin{definition}
    For a given state, the valence of the length $2m+1$ string in the $j$ sea
    is defined by $P_{m}^{(j)}=Q^{(j)}_{m,\max}-Q^{(j)}_{m,\min}+1$,
    that is:
\begin{equation}
P_{m}^{(j)} = \nu_{m}^{(j)} -\sum_{n\in\half\ZZ_{+}}\Big\{
\min(2m+1,2n+1)\, w_{n}^{(j)}\Big\}
+\sum_{l=1}^{L}\delta_{j,j_{l}}\,\min(2m+1, a_{\ell})
\end{equation}
where the representation at site $l$ is given by the $sl(N)$ weight
$(\underbrace{0,\ldots,0}_{j_{l}-1},a_{l},0,\ldots,0)$ and the
coefficients $w_{n}^{(j)}$ are given in equation (\ref{eq:defw}).
 \end{definition}
We can reformulate the spin (\ref{eq:spin-M}) of these states as
\begin{equation}
S_{j}= -\sum_{n\in\half\ZZ_{+}} (2n+1)\, w_{n}^{(j)}
+\sum_{l=1}^{L}\delta_{j,j_{l}}\,a_{l}
\label{eq:spin-w}
\end{equation}

We illustrate this definition with the previous three examples.

\paragraph{Examples}
\begin{enumerate}
\item For instance, the valences corresponding to a \slf are given by
\begin{eqnarray}
P_{m}^{(1)} &=& \nu_{m}^{(1)} -\sum_{n\in\half\ZZ_{+}}\Big\{
\min(2m+1,2n+1)\, w_{n}^{(1)}\Big\} +L\\
P_{m}^{(j)} &=& \nu_{m}^{(j)} -\sum_{n\in\half\ZZ_{+}}\Big\{
\min(2m+1,2n+1)\, w_{n}^{(j)}\Big\}\mb{for} 1<j\leq N
\end{eqnarray}
\item The valences for a \sls are 
 \begin{equation}
P_{m} = \nu_{m} -\sum_{n\in\half\ZZ_{+}}\,
\min(2m+1,2n+1)\, \nu_{n} +L\,\min(2m+1, s)
\end{equation}

\item The valences for an \alt are given by
\begin{eqnarray}
P_{m}^{(j_1)} &=& \nu_{m}^{(j_{1})} -\sum_{n\in\half\ZZ_{+}}\Big\{
\min(2m+1,2n+1)\, w_{n}^{(j_1)}\Big\} +\frac{L}{2}\,\min(2m+1,a_1)\\
P_{m}^{(j_2)} &=& \nu_{m}^{(j_{2})} -\sum_{n\in\half\ZZ_{+}}\Big\{
\min(2m+1,2n+1)\, w_{n}^{(j_2)}\Big\} +\frac{L}{2}\,\min(2m+1,a_2)\\
P_{m}^{(j)} &=& \nu_{m}^{(j)} -\sum_{n\in\half\ZZ_{+}}\Big\{
\min(2m+1,2n+1)\, w_{n}^{(j)}\Big\}\mb{for} j\neq j_{1},j_{2}
\end{eqnarray}
where we have supposed $j_{1}\neq j_{2}$.

When $j_{1}=j_{2}\equiv j_{0}$ (and $a_{1}\neq a_{2}$), the valences 
read
\begin{eqnarray*}
P_{m}^{(j_0)} &=& \nu_{m}^{(j_{0})} -\sum_{n\in\half\ZZ_{+}}\Big\{
\min(2m+1,2n+1)\, w_{n}^{(j_0)}\Big\} 
+\frac{L}{2}\,\Big(\min(2m+1,a_1)+\min(2m+1,a_2)\Big) \\
P_{m}^{(j)} &=& \nu_{m}^{(j)} -\sum_{n\in\half\ZZ_{+}}\Big\{
\min(2m+1,2n+1)\, w_{n}^{(j)}\Big\}\mb{for} j\neq j_{0}
\end{eqnarray*}
\end{enumerate}

\section{Vacuum states\label{sec:vacuum}}

In this section, we look for states that correspond to zero excitation
(no hole) configuration, i.e. 
\begin{equation}
P_{m}^{(j)} = \nu_{m}^{(j)}\,, \ \forall
j=1,\ldots,N-1 \mb{and} m\in\half\ZZ_{+}. 
\end{equation}
We call them \textit{vacuum states}. We have the following
theorem:
\begin{theorem}
\label{thm:vacuum}
We consider a spin chain based on $gl(N)$, with on each site $\ell$, a
representation given by the Young tableau fig. \ref{Youngtab},
characterized by $(a_{\ell},j_{\ell})$.\\
(i) For such a spin chain, the vacuum state is non-degenerate, and
 built on $\nu_{m}^{(j)}$ strings of length $(2m+1)$ in the $j$ sea, with
\begin{equation}
N\,\nu_{m}^{(j)} = \sum_{l=1}^{L}\delta_{2m+1,a_{l}}\, \min(j,j_{l})\,
\Big(N-\max(j,j_{l})\Big)
\label{string-vac}
\end{equation}
(ii) The vacuum state is antiferromagnetic, i.e. it is a spin 0 state under 
the $sl(N)$ symmetry algebra of the transfer matrix.
\end{theorem}
\prf
The vacuum states are defined by
$P_{m}^{(j)} = \nu_{m}^{(j)}$, $\forall
j=1,\ldots,N-1$ and $m\in\half\ZZ_{+}$, hence they must solve the
equations
\begin{eqnarray}
&&\sum_{n\in\half\ZZ_{+}}\Big\{\min(2m+1,2n+1)\, w_{n}^{(j)}\Big\}
=\sum_{l=1}^{L}\delta_{j,j_{l}}\,\,\min( a_{l},2m+1)
\label{eq.antif}\\
&& \forall j=1,\ldots,N-1 \mb{and} m\in\half\ZZ_{+} \nonumber
\end{eqnarray}
Performing the difference of two equations (\ref{eq.antif}), for $m$
and $m+\half$, and using the identities
\begin{eqnarray}
&&\min(a_{l},2m+1)=a_{l}\,\theta(2m\geq a_{l})+(2m+1)\,\theta(a_{l}>2m)
\\
&&\theta(2m+1\geq a_{l})-\theta(2m\geq a_{l}) = \delta_{2m+1,a_{l}}
\\
&&\theta(2m< a_{l})-\theta(2m+1< a_{l}) = \delta_{2m+1,a_{l}}
\end{eqnarray}
one gets
\begin{equation}
\sum_{n>m}\,w_{n}^{(j)}
=\sum_{l=1}^{L}\delta_{j,j_{l}}\,\theta(a_{l}>2m+1)\,.
\label{eq:w-aux}
\end{equation}
This proves that the system (\ref{eq.antif}) is triangular, and thus admits at most one solution, 
so that one has only to prove the existence of such a state. 

We take the values:
\begin{equation}
w_{m}^{(j)} =\sum_{l=1}^{L}\delta_{j,j_{l}}\,\delta_{a_{l},2m+1}
\label{eq.w-antif}
\end{equation}
Pluging these values into (\ref{eq.antif}), a direct
calculation
shows that these equations are all satisfied. Hence,
(\ref{eq.w-antif}) defines a vacuum state. 

Rewriting the expression
(\ref{eq.w-antif}) in terms of $\nu_{m}^{(j)}$ using (\ref{eq:defw}), we get the result
(\ref{string-vac}).

Using the values (\ref{eq.w-antif}) and the expression
(\ref{eq:spin-w}), it is then easy to compute that the $sl(N)$ eigenvalues 
of these states vanish.
\finprf

\begin{rmk}\label{rmk-Lpair}
Note that the relation (\ref{string-vac}) implies that 
$\sum_{l=1}^{L}\delta_{2m+1,a_{l}}\, \min(j,j_{l})\,(N-\max(j,j_{l}))$ must be a multiple of 
$N$. This constraint (on the existence of the vacuum state)
can be viewed as a constraint on the
length of the spin chain, with parameters $a_{l}$ and $j_{l}$, see
examples below. 
\end{rmk}

In words, the above theorem states that the vacuum state is built with
strings of length $a_{\ell}$ in each sea. The number of these strings 
in the sea $j$ is a 
function of $j$ and $j_{\ell}$,
$\ell=1,\ldots,L$. 

Even though the construction of Hamiltonian for a general choice of 
representation is difficult and not in the scope of this work, we may expect that for an approriate 
choice of the coupling constants the spin chain becomes a bipartite lattice ($L_0\geq 2$). 
In this case, the Marshall's theorem \cite{mar,limattis} can be applied and the state found in the 
theorem \ref{thm:vacuum} is the ground state for the corresponding Hamiltonian.

\paragraph{Examples}
\begin{enumerate}
\item For instance, the vacuum state corresponding to a \slf is given by
 \begin{equation}
 \nu_{0}^{(j)}=\frac{L(N-j)}{N}\mb{and} 
 \nu_{n}^{(j)}=0\,,\ 
\forall\, n\in\half\ZZ_{>0}\,,\qquad1\leq j\leq N-1\,.
\end{equation}
One recovers the usual antiferromagnetic ground state, with only real Bethe roots.
Their total number is $\frac{L(N-j)}{N}$, so that the
state exists only if $L(N-j)$ can be divided by $N$, i.e. $L$
multiple of $N$.

\item For a \sls we get
 \begin{equation}
\nu_{s-\half} = \frac{L}{2}\mb{and} \nu_{n} = 0\,,\ \forall\,
n\in\half\ZZ\,,\ n\neq s-\half\,.
\end{equation}
We get $\frac{L}{2}$ strings of length $2s$, as expected for the
ground state. One also recovers that $L$ must be even.

\item  If one considers an \alt\!\!, one gets for $n\in\half\ZZ$
 \begin{equation}
\nu_{n}^{(j)} = 0\mb{when}
n\neq n_{1}=\frac{a_{1}-1}{2}\mb{and}n\neq n_{2}=\frac{a_{2}-1}{2}\,.
\end{equation}
If one  supposes furthermore that $a_{1}\neq a_{2}$, 
\begin{equation}
 \nu_{n_{k}}^{(j)}=\left\{\begin{array}{l}\displaystyle
 \frac{Lj(N-j_{k})}{2N}\mb{for} j\leq j_{k}\\[2.1ex]
\displaystyle\frac{Lj_{k}(N-j)}{2N}\mb{for} j\geq j_{k}\end{array}\right.
 \qquad\ k=1,2
\end{equation}
\end{enumerate}

\section{Thermodynamical limit\label{sec:tdl}}

The Bethe equations cannot be solved in general however 
interesting features of the system can be obtained in the 
thermodynamical limit (i.e. $L\to\infty$). We will need the following 
definition:
\begin{definition}
A spin chain is called \emph{$L_{0}$-regular} $(L_{0}\in\ZZ_{+})$ when
the types of representations entering in its definition satisfy
\begin{equation}
j_{\ell+L_{0}}=j_{\ell} \mb{and} a_{\ell+L_{0}}=a_{\ell}\quad,\quad \forall 
\ell=1,\ldots L\;.
\end{equation}
The set of integers $\{\bar a_1,\ldots,\bar a_\cL\}$ corresponds to 
the distinct values in the set $\{a_\ell|1\leq\ell\leq L_0\}$ 
(let us stress that $\cL$ can be different from $L_{0}$).
Finally, we introduce
\begin{equation}
\cN=\{n_{\alpha}=\frac{\bar a_{\alpha}-1}{2}\;|\;1\leq\alpha\leq\cL\}\;.
\end{equation} 
\end{definition}
We also introduce the sets of indices defined by:
\begin{equation}
    \label{eq:set}
I_{\alpha}=\{j\in [1,L_{0}] \mb{s.t.} a_{j}=\bar a_{\alpha}\}\ ,\ \forall
\alpha\in[1,\cL]
\end{equation}
such that
\begin{equation}
\label{eq:s2s}
\sum_{\ell=1}^{L_{0}}\,(\ldots)_{\ell}\ =\
\sum_{\alpha=1}^{\cL}\,\sum_{\ell'\in I_{\alpha}}\,(\ldots)_{\ell'} \;.
\end{equation}
The cardinal $|I_{\alpha}|$ corresponds to the multiplicity of $\bar 
a_{\alpha}$ within a subset of $L_{0}$ sites.

\paragraph{Examples}
\begin{enumerate}
\item For a \slf\!\!, we recall that $a_{\ell}=j_{\ell}=1$ and thus one has
$L_{0}=1=\cL$,  $\cN=\{0\}$ and $I_{1}=\{1\}$.
\item For a \sls\!\!, we get $a_{\ell}=2s$, $j_{\ell}=1$ and 
one has
$L_{0}=1=\cL$,  $\cN=\{s-\half\}$ and $I_{1}=\{1\}$.
\item[3.]  For an \alt\!\!, one has $L_{0}=2=\cL$ if
$a_{1}\neq a_{2}$ (whatever the two values  $j_{1}$ and $j_{2}$
 are); while $L_{0}=2$ and $\cL=1$ if  $a_{1}= a_{2}$ and $j_{1}\neq 
j_{2}$. We have $\cN=\{\frac{a_{1}-1}{2},\frac{a_{2}-1}{2}\}$ and 
$I_{1}=\{1\}$, $I_{2}=\{2\}$ in the first case; 
$\cN=\{\frac{a_{1}-1}{2}\}$ and $I_{1}=\{1,2\}$ in the second case.
\end{enumerate}

{From} now on, we assume that the spin chain is $L_{0}$-regular.
Then, adding the constraint on the existence of a vacuum state (see
 remark \ref{rmk-Lpair}) to the above
regularity condition, one is led to take $L=p\,N\,L_{0}$, 
$p\in \ZZ_{+}$.

\subsection{Regularity and constraint on representations\label{sec:L0rep}}

The regularity hypothesis is also a natural 
assumption to solve the constraint (\ref{eq:CNSstring}) within the 
$L\to\infty$ limit. Indeed, this constraint applied to a regular spin 
chain is equivalent to 
\begin{equation}
\sum_{\ell=1}^{L_{0}}  
\left(\gamma^{(j)}_{\ell}+\frac{a^{(j)}_{\ell}}{2}\right)^{2p}\, 
=\,\sum_{\ell=1}^{L_{0}}  
\left(\gamma^{(j)}_{\ell}-\frac{a^{(j)}_{\ell}}{2}\right)^{2p}\,,
\quad 1\leq p\leq L_{0}
\label{eq:phaseL0}
\end{equation}
which is not affected by the limit. 

As already remarked, in the case of $gl(2)$,
the representations (\ref{eq:glN-weight}) describe all the
representations, so that there is in fact no constraint for this
algebra (whatever the value of $L_{0}$). Then, it is natural to wonder
if other kinds of representations can appear when $N>2$.

When $L_{0}=1$, a direct calculation proves that
the constraint (\ref{eq:phaseL0}) is equivalent to the 
equation (\ref{eq:phase}), so that the representations described by  
(\ref{eq:glN-weight}) are the only ones compatible with the string
hypothesis (whatever the value of $N$). 

For $L_{0}=2$, one can solve directly the 
equations (\ref{eq:phaseL0}).  One finds that $a^{(j)}_{\ell}\geq0$
and $\gamma^{(j)}_{\ell}$ must fulfil
one of the conditions (for $j=1,\ldots,N-1$):
\begin{equation}
\begin{cases}
 a^{(j)}_{1}\,\gamma^{(j)}_{1}=0 \mb{and} 
a^{(j)}_{2}\,\gamma^{(j)}_{2}=0\\
\mb{or}\\
a^{(j)}_{1}=2\,\eps^{(j)}\,\gamma^{(j)}_{2} \mb{and} 
a^{(j)}_{2}=-2\,\eps^{(j)}\,\gamma^{(j)}_{1}
\,,\quad\eps^{(j)}=\pm1\,,\\
\mb{or}\\
a^{(j)}_{1}=a^{(j)}_{2} \mb{and} 
\gamma^{(j)}_{2}=-\gamma^{(j)}_{1}\,.
\end{cases}
\end{equation}
Remark  that the first line (when applied for all $j$) corresponds to 
the representations we have studied in the present paper.
Plugging the value (\ref{eq:gamma}) of $\gamma^{(j)}_{\ell}$, one
gets
\begin{equation}
\begin{cases}
 a^{(j)}_{1}\,\Big(2\,\theta_1+\varpi^{(j)}_{1}+\varpi^{(j+1)}_{1}-j\Big)=0 
 \mb{and} 
a^{(j)}_{2}\,\Big(2\,\theta_2+\varpi^{(j)}_{2}+\varpi^{(j+1)}_{2}-j\Big)=0\,,\\
\mb{or}\\
2\,\theta_2 =
\eps^{(j)}\,a^{(j)}_{1}-(\varpi^{(j)}_{2}+\varpi^{(j+1)}_{2})+j 
\mb{and} 
2\,\theta_1 =
-\eps^{(j)}\,a^{(j)}_{2}-(\varpi^{(j)}_{1}+\varpi^{(j+1)}_{1})+j 
\,,\\
\mb{or}\\
a^{(j)}_{1}=a^{(j)}_{2} \mb{and} 
2\,(\theta_1+\theta_{2}) =
2\,j-(\varpi^{(j)}_{1}+\varpi^{(j+1)}_{1}+\varpi^{(j)}_{2}+\varpi^{(j+1)}_{2})
\,,
\end{cases}
\end{equation}
that one needs to
solve  in $\theta_{\ell}$, 
$a^{(j)}_{\ell}=\varpi^{(j)}_{\ell}-\varpi^{(j+1)}_{\ell}$ and
$\varpi^{(1)}_{\ell}$, with the conditions $a^{(j)}_{\ell}\geq0$.
This is still an intriguing problem, and we just give the complete
classification of solutions for $N=3$ (obtained by direct calculation). 

\begin{figure}[htb]
\begin{eqnarray}
    (c,b)\qquad
    \begin{picture}(60,30)
      \put(0,20){\line(1,0){10}}\put(20,20){\line(1,0){10}}\put(50,20){\line(1,0){10}}
      \put(0,10){\line(1,0){10}}\put(20,10){\line(1,0){10}}\put(50,10){\line(1,0){10}}
      \put(0,0){\line(1,0){10}}\put(20,0){\line(1,0){10}}
      \put(0,0){\line(0,1){20}}\put(10,0){\line(0,1){20}}
      \put(20,0){\line(0,1){20}}\put(30,0){\line(0,1){20}}
      \put(50,20){\line(0,-1){10}}\put(60,20){\line(0,-1){10}}
      \multiput(30,20)(4,0){8}{\line(1,0){2}}
      \multiput(30,10)(4,0){8}{\line(1,0){2}}
      \multiput(10,0)(4,0){3}{\line(1,0){2}}
      \multiput(10,10)(4,0){3}{\line(1,0){2}}
      \multiput(10,20)(4,0){3}{\line(1,0){2}}
      \put(2,-3){$\underbrace{~~~~~~~}_{\displaystyle b}$}
      \put(32,7){$\underbrace{~~~~~~~}_{\displaystyle c}$}
    \end{picture}
&&
\begin{picture}(60,30)
  \put(0,10){\line(1,0){20}}\put(50,10){\line(1,0){10}}
  \put(0,0){\line(1,0){20}}\put(50,0){\line(1,0){10}}
  \put(0,10){\line(0,-1){10}}\put(50,10){\line(0,-1){10}}
  \put(10,10){\line(0,-1){10}}\put(60,10){\line(0,-1){10}}
  \put(20,10){\line(0,-1){10}}
  \multiput(20,10)(4,0){8}{\line(1,0){2}}
  \multiput(20,0)(4,0){8}{\line(1,0){2}}
  \put(0,-3){$\underbrace{~~~~~~~~~~~~~~~}_{\displaystyle c}$}
\end{picture}\qquad(c,0)
\nonumber \\[4.2ex]
\begin{picture}(60,30)
\put(0,20){\line(1,0){10}}\put(20,20){\line(1,0){10}}\put(50,20){\line(1,0){10}}
\put(0,10){\line(1,0){10}}\put(20,10){\line(1,0){10}}\put(50,10){\line(1,0){10}}
\put(0,0){\line(1,0){10}}\put(20,0){\line(1,0){10}}
\put(0,0){\line(0,1){20}}\put(10,0){\line(0,1){20}}
\put(20,0){\line(0,1){20}}\put(30,0){\line(0,1){20}}
\put(50,20){\line(0,-1){10}}\put(60,20){\line(0,-1){10}}
\multiput(30,20)(4,0){8}{\line(1,0){2}}
\multiput(30,10)(4,0){8}{\line(1,0){2}}
\multiput(10,0)(4,0){3}{\line(1,0){2}}
\multiput(10,10)(4,0){3}{\line(1,0){2}}
\multiput(10,20)(4,0){3}{\line(1,0){2}}
\put(2,-3){$\underbrace{~~~~~~~}_{\displaystyle b}$}
\put(32,7){$\underbrace{~~~~~~~}_{\displaystyle c}$}
\end{picture}
&&
\begin{picture}(60,30)
  \put(0,10){\line(1,0){20}}\put(50,10){\line(1,0){10}}
  \put(0,0){\line(1,0){20}}\put(50,0){\line(1,0){10}}
  \put(0,10){\line(0,-1){10}}\put(50,10){\line(0,-1){10}}
  \put(10,10){\line(0,-1){10}}\put(60,10){\line(0,-1){10}}
  \put(20,10){\line(0,-1){10}}
  \multiput(20,10)(4,0){8}{\line(1,0){2}}
  \multiput(20,0)(4,0){8}{\line(1,0){2}}
  \put(0,-3){$\underbrace{~~~~~~~~~~~~~~~}_{\displaystyle b+c+1}$}
\end{picture}
\nonumber \\[5.2ex]
(b,c)\qquad\begin{picture}(60,30)
\put(0,20){\line(1,0){10}}\put(20,20){\line(1,0){10}}\put(50,20){\line(1,0){10}}
\put(0,10){\line(1,0){10}}\put(20,10){\line(1,0){10}}\put(50,10){\line(1,0){10}}
\put(0,0){\line(1,0){10}}\put(20,0){\line(1,0){10}}
\put(0,0){\line(0,1){20}}\put(10,0){\line(0,1){20}}
\put(20,0){\line(0,1){20}}\put(30,0){\line(0,1){20}}
\put(50,20){\line(0,-1){10}}\put(60,20){\line(0,-1){10}}
\multiput(30,20)(4,0){8}{\line(1,0){2}}
\multiput(30,10)(4,0){8}{\line(1,0){2}}
\multiput(10,0)(4,0){3}{\line(1,0){2}}
\multiput(10,10)(4,0){3}{\line(1,0){2}}
\multiput(10,20)(4,0){3}{\line(1,0){2}}
\put(2,-3){$\underbrace{~~~~~~~}_{\displaystyle c}$}
\put(32,7){$\underbrace{~~~~~~~}_{\displaystyle b}$}
\end{picture}
&&
\begin{picture}(60,30)
  \put(0,20){\line(1,0){20}}\put(50,20){\line(1,0){10}}
  \put(0,10){\line(1,0){20}}\put(50,10){\line(1,0){10}}
  \put(0,0){\line(1,0){20}}\put(50,0){\line(1,0){10}}
  \put(0,0){\line(0,1){20}}\put(50,0){\line(0,1){20}}
  \put(10,0){\line(0,1){20}}\put(60,0){\line(0,1){20}}
  \put(20,0){\line(0,1){20}}
  \multiput(20,20)(4,0){8}{\line(1,0){2}}
  \multiput(20,10)(4,0){8}{\line(1,0){2}}
  \multiput(20,0)(4,0){8}{\line(1,0){2}}
  \put(0,-3){$\underbrace{~~~~~~~~~~~~~~~}_{\displaystyle c}$}
\end{picture}\qquad (0,c)
\nonumber \\[4.2ex]
\begin{picture}(60,30)
\put(0,20){\line(1,0){10}}\put(20,20){\line(1,0){10}}\put(50,20){\line(1,0){10}}
\put(0,10){\line(1,0){10}}\put(20,10){\line(1,0){10}}\put(50,10){\line(1,0){10}}
\put(0,0){\line(1,0){10}}\put(20,0){\line(1,0){10}}
\put(0,0){\line(0,1){20}}\put(10,0){\line(0,1){20}}
\put(20,0){\line(0,1){20}}\put(30,0){\line(0,1){20}}
\put(50,20){\line(0,-1){10}}\put(60,20){\line(0,-1){10}}
\multiput(30,20)(4,0){8}{\line(1,0){2}}
\multiput(30,10)(4,0){8}{\line(1,0){2}}
\multiput(10,0)(4,0){3}{\line(1,0){2}}
\multiput(10,10)(4,0){3}{\line(1,0){2}}
\multiput(10,20)(4,0){3}{\line(1,0){2}}
\put(2,-3){$\underbrace{~~~~~~~}_{\displaystyle c}$}
\put(32,7){$\underbrace{~~~~~~~}_{\displaystyle b}$}
\end{picture}
&\qquad&
\begin{picture}(60,30)
  \put(0,20){\line(1,0){20}}\put(50,20){\line(1,0){10}}
  \put(0,10){\line(1,0){20}}\put(50,10){\line(1,0){10}}
  \put(0,0){\line(1,0){20}}\put(50,0){\line(1,0){10}}
  \put(0,0){\line(0,1){20}}\put(50,0){\line(0,1){20}}
  \put(10,0){\line(0,1){20}}\put(60,0){\line(0,1){20}}
  \put(20,0){\line(0,1){20}}
  \multiput(20,20)(4,0){8}{\line(1,0){2}}
  \multiput(20,10)(4,0){8}{\line(1,0){2}}
  \multiput(20,0)(4,0){8}{\line(1,0){2}}
\put(0,-3){$\underbrace{~~~~~~~~~~~~~~~}_{\displaystyle b+c+1}$}
\end{picture}
\nonumber
\end{eqnarray}
\caption{The remaining allowed representations for $L_{0}=2$ and $N=3$\label{fig:L2N3}}
\end{figure}
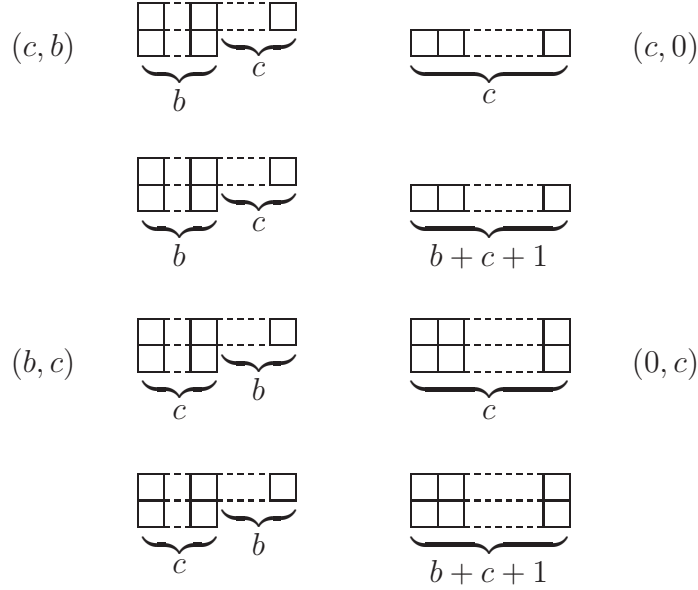

In the case of $sl(3)$, and discarding the case of arbitrary rectangular Young 
tableaux on each site, we get eight classes of solutions, which reduce
to four taking into account the symmetry between the two sites. 
In each case, one site is represented by a general Young tableau $(b,c)$, 
while the second one is rectangular  $(a,0)$ or $(0,a)$, but with $a$
determined by $b$ and $c$. We present in figure \ref{fig:L2N3} the
Young tableaux of the different cases.
As previously, the inhomogeneity parameters 
and the $gl(1)$ parameters are constrained:
\begin{equation}
\theta_1+\varpi_1^{(1)}=\half(b+1) +c+1
\mb{and} \theta_2+\varpi_2^{(1)}=\half(b+1)\,.
\end{equation}

\subsection{Limit of the Bethe equations for the vacuum}

For vacuum states, using the values (\ref{eq:Qminmax}) of $Q_{m,\min}^{(j)}$ and 
$Q_{m,\max}^{(j)}$ and the monotony 
of the $Q$'s, one gets
\begin{equation}
  Q_{m,k}^{(j)}=k-\half(\nu_{m}^{(j)}+1)\, \quad
k=1,\ldots,\nu_{m}^{(j)}.
\label{eq:Qmk-vide}
\end{equation}
Using the regularity hypothesis and equation (\ref{eq:s2s}), we simplify the  BAE 
(\ref{eq:bethe-argumt}) to:
\begin{eqnarray}
&&\sum_{n\in\cN}\left\{\sum_{\ell=1}^{\nu^{(j-1)}_{n}}
\Phi_{-1}^{(m,n)}(\lambda_{m,k}^{(j)}-\lambda_{n,\ell}^{(j-1)})
+\sum_{\atopn{\ell=1}{(n,\ell)\neq(m,k)}}^{\nu^{(j)}_{n}}
\Phi_{2}^{(m,n)}(\lambda_{m,k}^{(j)}-\lambda_{n,\ell}^{(j)})
+\sum_{\ell=1}^{\nu^{(j+1)}_{n}}
\Phi_{-1}^{(m,n)}(\lambda_{m,k}^{(j)}-\lambda_{n,\ell}^{(j+1)})
\right\}  \nonu
&& = -2\pi Q^{(j)}_{m,k}+\frac{L}{L_{0}}
\sum_{\alpha=1}^{\cL}\Big(\sum_{\ell\in I_\alpha}
\delta_{j,j_{\ell}}\Big)\,
\Phi_{\bar a_{\alpha}}^{(m)}(\lambda_{m,k}^{(j)})\\
&&\qquad \forall k=1,\ldots,\nu^{(j)}_{m}\ ,\ \forall
j=1,\ldots,N-1\ ,\ \forall m\in\half\ZZ_{+}\nonumber
\end{eqnarray}
	
Note that we can restrict ourself to the cases $m\in\cN$, keeping in
mind that $\nu_{m}^{(j)}=0$ when $m\not\in\cN$.

In the thermodynamical limit  $L\to\infty$, $Q_{m,k}^{(j)}/L$ becomes a 
continuous variable $x$, whose bounds are obtained from 
(\ref{eq:Qmk-vide}) and (\ref{string-vac}):
\begin{equation}
  -Y_{m}^{(j)}\leq x\leq Y_{m}^{(j)}
  \mb{with} Y_{m}^{(j)}=\frac{1}{2NL_{0}}\,\sum_{\ell=1}^{L_{0}}
\delta_{2m+1,a_{\ell}}\,\min(j,j_{\ell})\,\Big(N-\max(j,j_{\ell})\Big)\;.
\end{equation}
Then, the thermodynamical limit of the BAE reads
\begin{eqnarray}
&&\sum_{n\in\cN}\left\{
\int_{-Y_{n}^{(j-1)}}^{Y_{n}^{(j-1)}} dy\
\Phi_{-1}^{(m,n)}(\lambda_{m}^{(j)}(x)-\lambda_{n}^{(j-1)}(y))
+\int_{-Y_{n}^{(j)}}^{Y_{n}^{(j)}} dy\ 
\Phi_{2}^{(m,n)}(\lambda_{m}^{(j)}(x)-\lambda_{n}^{(j)}(y))
\right.\nonu
&&\qquad\left.
+\int_{-Y_{n}^{(j+1)}}^{Y_{n}^{(j+1)}} dy\
\Phi_{-1}^{(m,n)}(\lambda_{m}^{(j)}(x)-\lambda_{n}^{(j+1)}(y))\right\}
 \ =\  -2\pi\,x+
\frac{1}{L_{0}}\sum_{\alpha=1}^{\cL}\Big(\sum_{\ell\in I_\alpha}\delta_{j,j_{\ell}}\Big)\,
\Phi_{\bar a_{\alpha}}^{(m)}(\lambda_{m}^{(j)}(x))\qquad\qquad\\
&&\qquad \forall x\in [-Y_{m}^{(j)}\,,\,Y_{m}^{(j)}]\ ,\ \forall
j=1,\ldots,N-1\ ,\ \forall m\in\cN\nonumber
\end{eqnarray}
Making the change of variables $x\in [-Y_{m}^{(j)}\,,\,Y_{m}^{(j)}]
\to \lambda_{m}^{(j)}(x)\in]-\infty\,,\,\infty[$,  we get
\begin{eqnarray}
&&\sum_{n\in\cN}\left\{
\int_{-\infty}^{\infty} d\lambda \,\sigma_{n}^{(j-1)}(\lambda)\,
{\Phi_{-1}^{(m,n)}}(\lambda_{0}-\lambda)
+\int_{-\infty}^{\infty} d\lambda \,\sigma_{n}^{(j)}(\lambda)\, 
{\Phi_{2}^{(m,n)}}(\lambda_{0}-\lambda)
\right.\nonu
&&\qquad
\left.+\int_{-\infty}^{\infty} d\lambda \,\sigma_{n}^{(j+1)}(\lambda)\,
{\Phi_{-1}^{(m,n)}}(\lambda_{0}-\lambda)
\right\}\  =\
-2\pi\,x(\lambda_{0})+\frac{1}{L_{0}}\sum_{\alpha=1}^{\cL}\Big(\sum_{\ell\in I_\alpha}\delta_{j,j_{\ell}}\Big)\,
\Phi_{\bar a_{\alpha}}^{(m)}(\lambda_{0})\qquad\qquad\\
&&\qquad \forall \lambda_{0}\in\;
 ]-\infty\,,\,\infty[\ ,\ \forall
j=1,\ldots,N-1\ ,\ \forall m\in\cN\nonumber
\end{eqnarray}
where $\lambda_{0}$ stands for $\lambda_{m}^{(j)}(x)$, and 
$\sigma_{n}^{(j)}=dx/d\lambda_{n}^{(j)}$.

\subsection{Calculation of the densities}

To solve these equations, one performs a differentiation w.r.t.
$\lambda_{m}^{(j)}$ and a Fourier transform. One computes:
\begin{eqnarray}
\Psi^{(m,n)}_{-1}(\lambda) =\frac{d\Phi^{(m,n)}_{-1}(\lambda)}{d\lambda} &=& -2\,\sum_{s=\vert
m-n\vert}^{m+n} \frac{s+\half}{(s+\half)^2+\lambda^2}\\
\Psi^{(m,n)}_{2}(\lambda) = \frac{d\Phi^{(m,n)}_{2}(\lambda)}{d\lambda} &=& 2\,\left(
\frac{ m+n+1}{(m+n+1)^2+\lambda^2}+
\frac{\vert m-n\vert}{(m-n)^2+\lambda^2}+
2\sum_{s=\vert m-n\vert+1}^{m+n}
\frac{s}{s^2+\lambda^2}\right)\quad\\
\Psi^{(m)}_{a}(\lambda) = \frac{d\Phi^{(m)}_{a}(\lambda)}{d\lambda} &=& 2\,
\sum_{s=|\frac{a}{2}- m|+1}^{\frac{a}{2}+m} \frac{s}{s^2+\lambda^2}
+2\;\theta(\frac{a}{2}> m)\;\frac{\frac{a}{2}-m}{(\frac{a}{2}-m)^2+\lambda^2}
\end{eqnarray}
We normalize the Fourier transform as:
\begin{equation}
 \wh{f}(p)  = \frac{1}{2\pi}\int_{-\infty}^{\infty}e^{ip\lambda} \;
 f(\lambda)\; d\lambda\,. 
 \end{equation}
Explicitly, one finds:
\begin{eqnarray}
\wh\Psi^{(m,n)}_{-1}(p) &=&
-\,\exp\Big(-|p|\big(\max(m,n)+\half\big)\Big)
\;\frac{\sinh\Big(|p|\big(\min(m,n)+\half\big)\Big)}{\sinh\frac{|p|}{2}}
\\
\wh\Psi^{(m,n)}_{2}(p) &=& 
-2\,\cosh\frac{|p|}{2}\;\wh\Psi^{(m,n)}_{-1}(p) - \delta_{m,n}\qquad
\\
\wh\Psi^{(m)}_{a}(p) &=& \exp\Big(-|p|\max(\frac{a}{2},m+\half)\Big)
\;\frac{\sinh\Big(|p|(\min(\frac{a}{2},m+\half)\Big)}{\sinh\frac{|p|}{2}}
\end{eqnarray}
We introduce $\wh\Psi(p)$,
a $(N-1)\times (N-1)$  matrix of $\cL\times \cL$ blocks:
\begin{equation}
\wh\Psi(p)=\left(\begin{array}{cccc}
\II+\wh\Psi_{2}(p) & \wh\Psi_{-1}(p) &0 & 0\\
\wh\Psi_{-1}(p) & \II+\wh\Psi_{2}(p) & \wh\Psi_{-1}(p)&0 \\
 & \ddots &\ddots & \\
0 & \wh\Psi_{-1}(p) & \II+\wh\Psi_{2}(p) &\wh\Psi_{-1}(p) \\
0 & 0 & \wh\Psi_{-1}(p) & \II+\wh\Psi_{2}(p)
\end{array}\right)
\end{equation}
where the blocks are  defined by
\begin{equation}
\label{eq:hatpsi}
\wh\Psi_{j}(p)=\left(\begin{array}{cccc}
\wh\Psi^{(n_{1},n_{1})}_{j}(p) &\wh\Psi^{(n_{1},n_{2})}_{j}(p) &
\ldots &\wh\Psi^{(n_{1},n_{\cL})}_{j}(p)\\
\wh\Psi^{(n_{2},n_{1})}_{j}(p) &\wh\Psi^{(n_{2},n_{2})}_{j}(p) &
\ldots &\wh\Psi^{(n_{2},n_{\cL})}_{j}(p)\\
\vdots & \ddots &\ddots &\vdots\\
\wh\Psi^{(n_{\cL},n_{1})}_{j}(p) &\wh\Psi^{(n_{\cL},n_{2})}_{j}(p) &
\ldots &\wh\Psi^{(n_{\cL},n_{\cL})}_{j}(p)\\
\end{array}\right)\,.
\end{equation}
In the same way, the BAE's r.h.s. becomes
\begin{equation}
\Lambda(p)=\frac{1}{L_{0}}\sum_{\alpha=1}^{\cL}\ \sum_{\ell\in I_\alpha}\left(\begin{array}{c}
\delta_{1,j_{\ell}} \\ \delta_{2,j_{\ell}}  \\ \vdots \\ \delta_{N-1,j_{\ell}}
\end{array}\right)\,\otimes\,\left(\begin{array}{c}
\wh\Psi^{(n_{1})}_{\bar a_{\alpha}}(p)\\ \wh\Psi^{(n_{2})}_{\bar a_{\alpha}}(p) \\
\vdots \\ \wh\Psi^{(n_{\cL})}_{\bar a_{\alpha}}(p)
\end{array}\right)
\end{equation}
and the (Fourier transformed) unknown variables  $\wh\sigma_{n}^{(j)}(p)$
are gathered in the column vector
\begin{equation}
\label{eq:sigma}
\wh\Sigma(p)=\left(\begin{array}{c}
\wh\Sigma_{1}(p) \\ \wh\Sigma_{2}(p) \\ \vdots \\ \wh\Sigma_{N-1}(p)
\end{array}\right)
\mb{with}
\wh\Sigma_{j}(p)=\left(\begin{array}{c}
\wh \sigma_{n_{1}}^{(j)}(p) \\ \wh \sigma_{n_{2}}^{(j)}(p) \\
\vdots \\ \wh \sigma_{n_{\cL}}^{(j)}(p)
\end{array}\right)\,.
\end{equation}

Finally, one is led to the following form of the BAE:
\begin{equation}
2\pi\,\wh\Psi(p)\,\wh\Sigma(p)=\Lambda(p)
\label{BAE-TF}
\end{equation}
Using the explicit forms of the functions, the matrix $\wh\Psi(p)$ 
reduces to 
\begin{equation}
  \wh\Psi(p)=\cA(p) \otimes \big(-\wh\Psi_{-1}(p)\big)
 \;,\quad\cA(p)=\left(\begin{array}{ccccc}
2\cosh(\frac{\vert p\vert}{2}) & -1 & 0 & \ldots & 0\\
-1 & 2\cosh(\frac{\vert p\vert}{2}) & -1 & \ddots& \vdots \\
0 & \ddots &\ddots & \ddots & 0\\
\vdots & \ddots & \ \ -1\ \  & 2\cosh(\frac{\vert p\vert}{2}) & -1 \\
 0 & \ldots & 0 & -1 & 2\cosh(\frac{\vert p\vert}{2})
\end{array}\right)
\end{equation}
with $\cA(p)$ a $(N-1)\times(N-1)$ matrix, reminiscent of the $sl(N)$ 
Cartan matrix. Let us remark that the matrix $\cA(p)$ is independant 
of the type of representations.

{{From}} this equation, one can deduce the densities 
$\wh \sigma_{n}^{(j)}(p)$ by inverting the matrix $\wh\Psi(p)$:
\begin{equation}
\wh\Sigma(p)=\frac{1}{2\pi}\,\Big(\cA(p)^{-1}
\otimes \big(-\wh\Psi_{-1}(p)\big)^{-1}\Big)\;\Lambda(p)
\label{BAE-TF-inv}
\end{equation}
where the inverse of the matrix $\cA(p)$ is given by \cite{sutherland}
\begin{equation}
(\cA(p)^{-1})_{ij}=
\frac{\displaystyle\sinh\Big((N-\max(i,j))\frac{\vert p\vert}{2}\Big) \;
\sinh\Big(\min(i,j)\frac{\vert p\vert}{2}\Big)} 
{\displaystyle\sinh\Big(N\frac{\vert p\vert}{2}\Big) \; 
\sinh\Big(\frac{\vert p\vert}{2}\Big)}\;.
\label{eq:psi}
\end{equation}
The previous computations allow us to find a compact expression for the 
densities given in the following theorem:
\begin{theorem} 
\label{thm:density}
Let us consider a $L_{0}$-regular spin chain based on $gl(N)$. 
For the vacuum, the only non vanishing densities
are, for $1\le k \le N-1$
and $n_{\alpha}\in \cN$ $(1\leq \alpha \leq \cL)$,
\begin{equation}
\sigma^{(k)}_{n_\alpha}(\lambda)=\frac{1}{N L_{0}}\sum_{\ell\in I_{\alpha}} 
 \ \ \sum_{q=(\vert k-j_\ell\vert+1)/2}^{(k+j_\ell-1)/2} \quad
\frac{\displaystyle \sin\Big(\frac{2q\pi}{N}\Big)}
{\displaystyle \cosh\Big(\frac{2\pi}{N}\lambda\Big) - 
\cos\Big(\frac{2q\pi}{N}\Big)}
\label{eq:densitemagique}
\end{equation}
where the sets $I_{\alpha}$ have been introduced in (\ref{eq:set}).

For the particular case of $gl(2)$ spin chains, the non-vanishing 
densities further simplify to:
\begin{equation}
\sigma_{s_\alpha-\half}(\lambda)=\frac{|I_{\alpha}|}{2 L_{0}}\;
\frac{1}{\displaystyle \cosh\pi\lambda} ,\quad \alpha=1,\ldots,\cL
\label{eq:densitemagique-gl2}
\end{equation}
where $s_{\alpha}=n_{\alpha}+\half$ denote the spins on the chain, 
and $|I_{\alpha}|$ is the cardinal of $I_{\alpha}$.
\end{theorem}
\prf We deduce from relation (\ref{BAE-TF-inv}) the
following expression for the vectors $\wh\Sigma_i(p)$ defined in
(\ref{eq:sigma}):
\begin{eqnarray}
\label{eq:Sigmath}
    \displaystyle
\wh\Sigma_i(p) &=&\frac{1}{2\pi\,L_0}
\sum_{\alpha=1}^{\cL}\ \Big(\sum_{\ell\in I_\alpha} (\cA(p)^{-1})_{i,j_\ell}\Big)\;
\big(-\wh\Psi_{-1}(p)\big)^{-1}\;\Lambda_{\alpha}(p)
\end{eqnarray}
\mbox{with}
\begin{eqnarray}
&& \Lambda_{\alpha}(p)=
\left(\begin{array}{c}
\displaystyle
\exp\Big(-\frac{|p|}{2}\max(\bar a_\alpha,\bar a_1)\Big)
\;\frac{\sinh\Big(\frac{|p|}{2}(\min(\bar a_\alpha,\bar a_1)\Big)}{\sinh\frac{|p|}{2}}\\
\vdots \\ 
\displaystyle
\exp\Big(-\frac{|p|}{2}\max(\bar a_\alpha,\bar a_\cL)\Big)
\;\frac{\sinh\Big(\frac{|p|}{2}(\min(\bar a_\alpha,\bar a_\cL)\Big)}{\sinh\frac{|p|}{2}}
\end{array}\right)
\label{eq:hS}
\end{eqnarray}
Remarking that the action of the matrix $-\wh\Psi_{-1}(p)$
on the elementary vector $e_{\alpha}\in \CC^\cL$ (with 1 in position
$\alpha$ and 0 elsewhere) gives the vector $\Lambda_{\alpha}(p)$, we
deduce that
$$\big(-\wh\Psi_{-1}(p)\big)^{-1}\;\Lambda_{\alpha}(p)=e_{\alpha}\;,\ 
\forall\ \alpha=1,\ldots\cL\;.
$$ 
Therefore, the projection on the elementary vector of relation (\ref{eq:Sigmath}) gives
\begin{equation}
\wh\sigma^{(i)}_{n_\alpha}(p)=\frac{1}{2\pi L_{0}}\sum_{\ell\in I_{\alpha}} 
\frac{\displaystyle\sinh\Big((N-\max(i,j_{\ell}))\frac{\vert p\vert}{2}\Big) \;
\sinh\Big(\min(i,j_{\ell})\frac{\vert p\vert}{2}\Big)} 
{\displaystyle\sinh\Big(N\frac{\vert p\vert}{2}\Big) \; 
\sinh\Big(\frac{\vert p\vert}{2}\Big)}\;.
\label{eq:hatsigmas}
\end{equation}
 Computing the inverse  Fourier
transform with the method of residues, and  using the trigonometric linearization formula
$$\displaystyle \frac{\sin ax}{\sin x} = \sum_{q=-(a-1)/2}^{(a-1)/2} \cos
2qx\;,$$ one is led to the equation (\ref{eq:densitemagique}).

The expression (\ref{eq:densitemagique-gl2}) is then straightforward 
when focusing on then $gl(2)$ case.
\finprf


\subsection{Examples}
 
\begin{figure}[htb]
\begin{center}
\epsfig{file=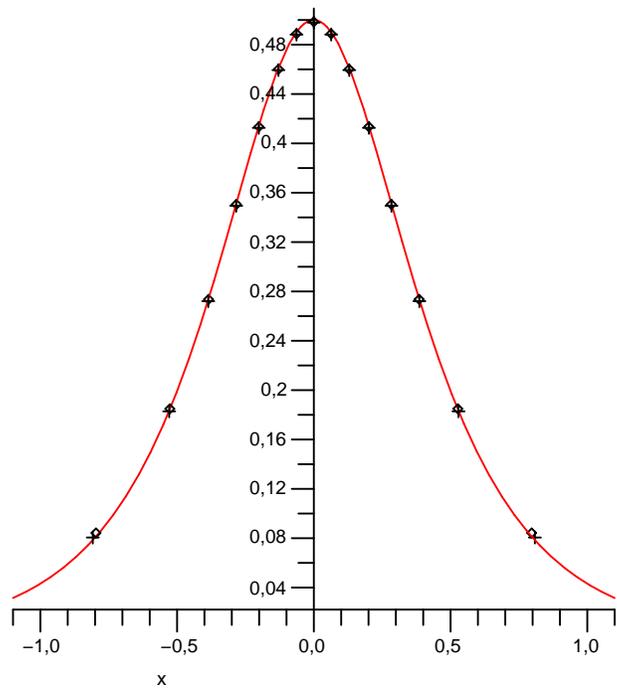,width=10cm}
\end{center}
\caption{Numerical vs analytical densities for $gl(2)$ 
alternating $(s=\half,s=\frac{3}{2})$ spin chain 
\label{fig:numeric}}
\end{figure}

\begin{figure}[htb]
\begin{center}
\epsfig{file=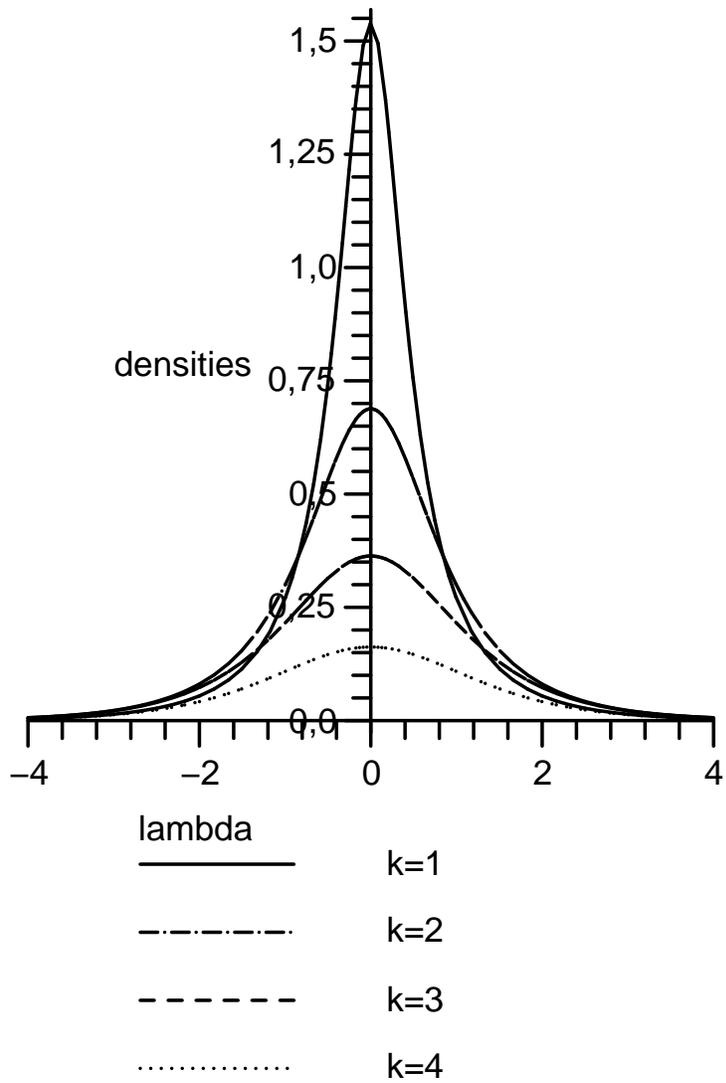,width=15cm}
\end{center}
\caption{Vacuum state densities for $gl(5)$ spin chains with fundamental
representations \label{fig:dens-fond}}
\end{figure}

\begin{figure}[htb]
\begin{center}
\epsfig{file=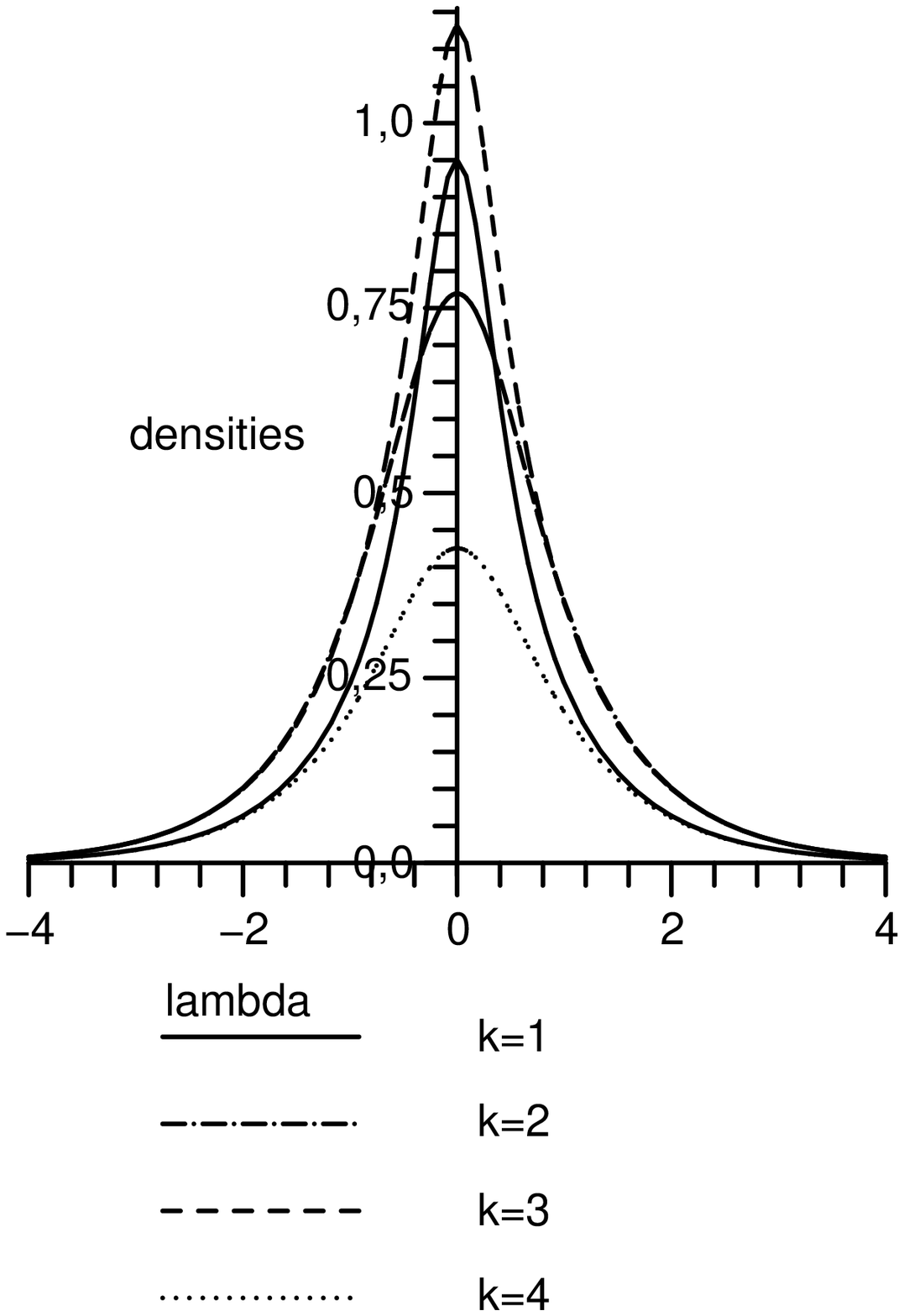,width=15cm}
\end{center}
\caption{Vacuum state densities for $gl(5)$ alternating $(a,1)-(a,3)$ spin chains 
\label{fig:dens-altern}} 
\end{figure}

In this subsection, we treat the thermodynamic limit for different examples:
\begin{enumerate}
\item
For instance, equation (\ref{eq:hatsigmas}) corresponding to a \slf simplifies to:
\begin{equation}
\wh\sigma_{0}^{(k)}(p) = \frac{\sinh\big((N-k)\frac{\vert p\vert}{2}\big)} 
{2\pi\,\sinh\big(N\frac{\vert p\vert}{2}\big)} \qquad 1 \le k \le N-1
\end{equation}
and the real roots densities $\sigma_{0}^{(k)}(\lambda)$ are given by, 
see (\ref{eq:densitemagique}), in accordance with \cite{sutherland,KulResh}
\begin{equation}
\sigma_{0}^{(k)}(\lambda) = \frac{1}{N}\;
\frac{\sin\big(\frac{\pi}{N}k\big)} 
{\cosh\big(\frac{2\pi}{N}\lambda\big) - \cos\big(\frac{\pi}{N}k\big)}
\label{eq:dens-fond-glN}
\end{equation}
This result is illustrated in figure \ref{fig:dens-fond} where the 
densities (\ref{eq:dens-fond-glN})
are plotted for $N=5$.
\item
For a \sls\!\!, one obtains
 the only non-vanishing density:
\begin{equation}
  \sigma_{s-\half}(\lambda) =\frac{1}{2\,\cosh(\pi\lambda)}\;.
  \label{eq:dens-s}
\end{equation}
\item[3-A.]
If one considers an \alt\!\!, with $a_{1}\neq a_{2}$, the non-vanishing 
densities take the following form (we remind that
$n_{\alpha}=\half(a_{\alpha}-1)$), for $k=1,\ldots N-1$,
\begin{equation}
\sigma^{(k)}_{n_\alpha}(\lambda)=\frac{1}{2N} 
\sum_{q=(\vert k-j_\alpha\vert+1)/2}^{(k+j_\alpha-1)/2} \quad
\frac{\displaystyle \sin\Big(\frac{2q\pi}{N}\Big)}
{\displaystyle \cosh\Big(\frac{2\pi}{N}\lambda\Big) - 
\cos\Big(\frac{2q\pi}{N}\Big)}\;,\ \alpha=1,2
\end{equation}
in accordance with the results of \cite{martins}.

For the particular case of an alternating $sl(2)$ spin chain 
with spin $s_1$ and $s_2$, the two non-vanishing densities
simplify to:
\begin{equation}
\sigma_{s_1-\half}(\lambda)=\sigma_{s_2-\half}(\lambda)
=
\frac{1}
{4\displaystyle \cosh(\pi\lambda)}\;.
\end{equation}
We remark that the densities do not depend on the spin, i.e. they have
the same expression, whatever the length of the string is. Their curve is
plotted in figure \ref{fig:numeric}, where, as a by-product, the comparison with the numerical
solutions confirms their independance from the length of the string
(see also footnote \ref{foot:length}). This fact has been noticed in 
\cite{dewo} for an alternating XXZ spin-$\half$, spin-$1$ chain. 
\item[3-B.]
If one considers an \alt\!\!, with $a_{1}= a_{2}$ and
$j_{1}\neq j_{2}$, the non-vanishing 
densities read (with $n=\half(a_{1}-1)$ and $k=1,\ldots,N-1$)
\begin{eqnarray}
\sigma^{(k)}_{n}(\lambda) &=&\frac{1}{2N}\ \sum_{\ell=1}^2\ 
\sum_{q=(\vert k-j_\ell\vert+1)/2}^{(k+j_\ell-1)/2} \quad
\frac{\displaystyle \sin\Big(\frac{2q\pi}{N}\Big)}
{\displaystyle \cosh\Big(\frac{2\pi}{N}\lambda\Big) - 
\cos\Big(\frac{2q\pi}{N}\Big)}\;.
\label{eq:boubou}
\end{eqnarray}
In figure \ref{fig:dens-altern}, we plot the above densities for $j_1=1$ 
and $j_2=3$ with $N=5$. There, we did not specify the values $a_{1} = 
a_{2}$, since, as above, the expression (\ref{eq:boubou})
is independent of them.
\end{enumerate} 

\subsection{Spin chains with periodic array of impurities\label{sec:periodic-array}}

We consider a $gl(N)$ spin chain in fundamental representations 
on each site except for the sites $p L_0$ ($1\leq p\leq L/L_0$) 
which carry a representation with the Young tableau characterized by $(j,a)$.
These sites can be interpreted as periodic impurities spread along the chain
\cite{fuka}.
In our framework, this chain is a particular choice of a $L_0$-regular chain with
\begin{equation}
\cN=\{0,\frac{a-1}{2}\}\mb{,}I_1=\{1,\dots,L_0-1\}\mb{and}I_2=\{L_0\}\;.
\end{equation}
We will assume that $L_{0}\geq3$, the cases $L_{0}=1,2$ have been 
already treated in the previous examples.

In this case, a local integrable Hamiltonian can be easily found amongst the 
conserved quantities in the expansion of the transfer matrix (\ref{transf}).
We fix the $gl(1)$ eigenvalue demanding that $\varpi_{\ell}^{(1)}$ is such that
$\theta_{\ell}=0$, $\forall\ell$. Then, the Hamiltonian
 can be written explicitly as
\begin{eqnarray}
H=\frac{d}{d\lambda}\ln(t(\lambda))\Big|_{\lambda=0}\propto
 \sum_{\atopn{n=2}{n,n-1\not\in L_0\ZZ_+}}^L \cP_{n-1,n} 
+~\sum_{\atopn{p=1}{p\in L_0\ZZ_+}}^{L}\big(~\cP_{p-1,p+1}~\cL_{p+1,p}(0)
+i~\big)~\cL_{p-1,p}^{-1}(0)
\end{eqnarray}
where the site $0$ is identified with $L$ and $\cL_{a,p}(\lambda)=\sum_{j,k}
E_{jk}\otimes (\lambda+i \cE_{kj})$. The matrices $\cE_{jk}$ are the
generators of $gl(\enne)$ in the representation $(j,a)$.

The energy per site is 
\begin{equation}
E=\frac{1}{iL}\frac{\Lambda'(0)}{\Lambda(0)}
\end{equation}
where $\Lambda(\lambda)$ is given by (\ref{ClosedEigen}) 
and the Drinfeld polynomials are computed using the formulas (\ref{drinP}),
(\ref{eq:glN-weight}) and (\ref{eq:inhomg}):
\begin{eqnarray}
P_1(\lambda)&=&(\lambda+i)^{L-\frac{L}{L_0}}(\lambda+i\;\frac{j+a}{2})^{\frac{L}{L_0}}\\
P_k(\lambda)&=&\lambda^{L-\frac{L}{L_0}}\times
\begin{cases}\displaystyle (\lambda+i\;\frac{j+a}{2})^{\frac{L}{L_0}}\mb{for}1<k\leq j\\
\displaystyle(\lambda+i\;\frac{j-a}{2})^{\frac{L}{L_0}}\mb{for}k> j
\end{cases}
\end{eqnarray}
The factor $i$ in the definition of the energy allows us to obtain a real energy.
 
\begin{figure}[htb]
\begin{center}
\epsfig{file=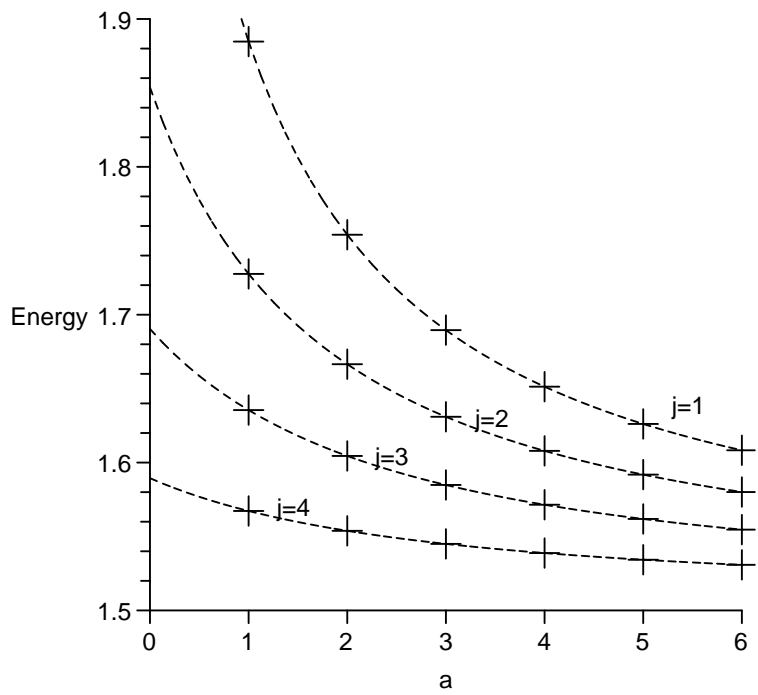,width=10cm}
\end{center}
\caption{Energy $E-\cC$ in terms of $a$ and $j$ (for $L_0=5$, $N=5$)\label{fig:nrj}}
\end{figure}

For the vacuum, the energy per site can be computed using the string hypothesis
\begin{equation}
E=\cC+\frac{1}{L}\sum_{n=1}^{\nu_0^{(1)}}\frac{1}{(\lambda_{0,n}^{(1)})^2+\frac{1}{4}}+
\frac{1}{L}\sum_{n=1}^{\nu_{\frac{a-1}{2}}^{(1)}}
\frac{a}{(\lambda_{\frac{a-1}{2},n}^{(1)})^2+\frac{a^2}{4}}
\end{equation}
where $\cC=-1+\frac{1}{L_0}-\frac{2}{L_0(j+a)}$ and the multiplicities
$\nu^{(1)}_{n}$ are given in (\ref{string-vac}). In the thermodynamical limit, we get
\begin{equation}
E-\cC=\int_{-\infty}^{\infty}d\lambda \Big(
\sigma_0^{(1)}(\lambda)\;\frac{1}{\lambda^2+\frac{1}{4}}\;
+\;\sigma_{\frac{a-1}{2}}^{(1)}(\lambda)\; \frac{a}{\lambda^2+\frac{a^2}{4}}\Big)
\end{equation}
where the densities are given by
\begin{equation}
\sigma^{(1)}_{0}(\lambda)=\frac{L_{0}-1}{NL_{0}} \ 
\frac{\displaystyle \sin\Big(\frac{\pi}{N}\Big)}
{\displaystyle \cosh\Big(\frac{2\pi}{N}\lambda\Big) -
\cos\Big(\frac{\pi}{N}\Big)} 
\;,\ 
\sigma^{(1)}_{(a-1)/2}(\lambda)=\frac{1}{NL_{0}}\ 
\frac{\displaystyle \sin\Big(\frac{j\pi}{N}\Big)}
{\displaystyle \cosh\Big(\frac{2\pi}{N}\lambda\Big) -
\cos\Big(\frac{j\pi}{N}\Big)}\;.
\end{equation}
Following the lines of \cite{sutherland}, we transform this relation as follows 
(using Plancherel's theorem)
\begin{equation}
E-\cC=\frac{L_{0}-1}{L_{0}}\int_{-\infty}^{\infty}dx\;e^{-|x|/2}
\frac{\sinh((N-1)x/2)}{\sinh(N\,x/2)}
+\frac{1}{L_{0}}\int_{-\infty}^{\infty}dx\;e^{-a|x|/2}
\frac{\sinh((N-j)x/2)}{\sinh(N\,x/2)}
\end{equation}
Note that the integrand is even. We define the new variable $y=e^{-Nx}$
which allows us to write
\begin{equation}
E-\cC=2\,\frac{L_{0}-1}{NL_{0}}\int_{0}^{1}dy\;\frac{y^{1/N-1}-1}{1-y}
+\frac{2}{NL_{0}}\int_{0}^{1}dy\;
\frac{ y^{\frac{a+j}{2N}-1} - y^{\frac{a-j}{2N}} }{1-y}
\end{equation}
Using the following formula (see e.g. \cite{grazy})
\begin{equation}
\int_0^1dy\;\frac{y^{\mu-1}-y^{\nu-1}}{1-y}=\psi(\nu)-\psi(\mu)\;,
\end{equation}
we obtain finally
\begin{equation}
E-\cC=\frac{2}{N}\Big(\psi(1)-\psi(\frac{1}{N})\Big)+
\frac{2}{NL_0}\Big(\psi(\frac{a-j}{2N}+1)-\psi(\frac{a+j}{2N}) 
-\psi(1)+\psi(\frac{1}{N})\Big)
\end{equation}
where $\psi(x)=\frac{d}{dx}\ln\Gamma(x)$ is the Euler digamma function. 
The first term corresponds to the case of a chain without impurity
($a=j=1$), as computed in \cite{sutherland}, while the second term is the
correction due to the impurities. \\
On figure \ref{fig:nrj}, we represent the energy per site, $E-\cC$, for
different types of impurities characterized by $a$ and $j$ (with $L_0=5$
and $N=5$).

\section*{Acknowledgments}
This work is supported by the TMR Network 
`EUCLID. Integrable models and applications: from strings to condensed
matter', contract number HPRN-CT-2002-00325.\\
ER thanks the Theoretische Physik group at Bergische Universit{\"a}t Wuppertal,
and specially F. Goehmann, for fruitful and stimulating discussions.\\
NC thanks Institut Universitaire de France and LAPTH.

\end{document}